\RequirePackage{fix-cm}
\documentclass[smallcondensed]{svjour3}     
\smartqed  

\usepackage{cite}
\usepackage{amsmath,amssymb,amsfonts}
\usepackage{algorithmic}
\usepackage{graphicx}
\usepackage{textcomp}
\usepackage[table,dvipsnames]{xcolor}
\usepackage{hyperref}
\usepackage{fancybox,framed}
\usepackage{multirow,blindtext}
\usepackage{tabularx}
\usepackage{xurl}
\usepackage{ifthen}
\usepackage[utf8]{inputenc}
\usepackage{booktabs}
\newcommand{\mycomment}[1]{}
\usepackage{pdflscape}
\usepackage[round]{natbib}

\journalname{Empirical Software Engineering}
\begin{document}
\title{Exploring Security Practices in Infrastructure as Code: An Empirical Study}


\author{Alexandre Verdet \and
        Mohammad Hamdaqa \and
        Leuson Da Silva \and
        Foutse Khomh
}


\institute{Alexandre Verdet \at
              Polytechnique Montreal, Montreal, Canada\\
              \email{alexandre.verdet@polymtl.ca}           
           \and
           Mohammad Hamdaqa \at
              Polytechnique Montreal, Montreal, Canada\\
              \email{mhamdaqa@polymtl.ca}           
           \and
           Leuson Da Silva \at
              Polytechnique Montreal, Montreal, Canada\\
              \email{leuson-mario-pedro.da-silva@polymtl.ca}           
           \and
           Foutse Khomh \at
              Polytechnique Montreal, Montreal, Canada\\
              \email{foutse.khomh@polymtl.ca}           
}

\date{Received: date / Accepted: date}


\maketitle

\begin{abstract}
\label{abstract}
Cloud computing has become popular thanks to the widespread use of Infrastructure as Code (IaC) tools, allowing the community to conveniently manage and configure cloud infrastructure using scripts. However, the scripting process itself does not automatically prevent practitioners from introducing misconfigurations, vulnerabilities, or privacy risks. As a result, ensuring security relies on practitioners’ understanding and the adoption of explicit policies, guidelines, or best practices. In order to understand how practitioners deal with this problem, in this work, we perform an empirical study analyzing the adoption of IaC scripted security best practices. First, we select and categorize widely recognized Terraform security practices promulgated in the industry for popular cloud providers such as AWS, Azure, and Google Cloud. Next, we assess the adoption of these practices by each cloud provider, analyzing a sample of 812 open-source projects hosted on GitHub. For that, we scan each project’s configuration files, looking for policy implementation through static analysis (checkov). Additionally, we investigate GitHub measures that might be correlated with adopting these best practices. The category \textit{Access policy} emerges as the most widely adopted in all providers, while \textit{Encryption in rest} are the most neglected policies. Regarding GitHub measures correlated with best practice adoption, we observe a positive, strong correlation between a repository number of stars and adopting practices in its cloud infrastructure. Based on our findings, we provide guidelines for cloud practitioners to limit infrastructure vulnerability and discuss further aspects associated with policies that have yet to be extensively embraced within the industry.
\keywords{Security Vulnerabilities \and Infrastructure as Code \and Policy Misconfiguration}
\end{abstract}

\section{Introduction}
\label{introduction}
Infrastructure as Code (IaC) makes it possible to deploy large cloud infrastructures while retaining the benefits of software development, such as code reuse, collaboration, testing, and static code analysis. 
With DevOps and continuous integration, IaC scripts have been integrated into production pipelines to further improve automation~\citep{spinellis, humble2010continuous}.

Over time, practitioners and researchers have been interested in IT security and privacy, considering the rise of new user concerns following major data leaks and surveillance.
In order to address those new related challenges, several data residency regulations have been introduced, such as the GDPR, CCPA, PIPEDA, HIPAA \cite{european_commission_regulation_2016, ccpa, pipeda}. 
Knowing that more systems are moved and/or deployed to the cloud, there is a growing demand for secure IaC solutions. 
In order to help practitioners comply with specific regulations, organizations and consortiums establish lists of security best practices. With time, some of those lists gained recognition among practitioners, becoming industry standards.

This way, security guidelines from those regulations establish sets of best practices, which actual implementations can be called policies.
While IaC tools such as Terraform \cite{terraform}, AWS CloudFormation \cite{cloudformation}, and Azure Resource Manager (ARM) \cite{azure_resource_manager} simplify software infrastructure provisioning, they shift responsibility for security risks to operational teams \citep{sharma}. 
However, writing security policies as code can be challenging. 
For instance, improper configuration can compromise security and put sensitive cloud data at risk \citep{sengupta}. 
In order to mitigate or avoid these issues, new standards have been proposed that IaC configurations must meet \citep{KEMP2018928}.
However, implementing these standards remains uneven because not all compliance guidelines can be translated into configuration implementation policies (actual identifiable code patterns), and not all policies are uniformly enforced.

As a result, the responsibility of infrastructure security is shared between practitioners and cloud providers and is often targeted by data protection regulations like GDPR or CCPA. 
Unsecured infrastructure can lead to unauthorized access to data or server instances, compromising the overall system.
To understand the usage and associated challenges with IaC, Guerriero et al. (\citeyear{guerriero2019adoption}) perform a qualitative study interviewing practitioners. 
The authors report that maintaining IaC code is one of the current challenges, while Terraform was observed as the most popular infrastructure provisioning tool. 
Regarding the usage of Terraform, GitHub currently indexes 109K Terraform files with AWS configurations, compared to 35.7K with Azure and 51.5K with Google Cloud. 
In the same way, Iosif et al. (\citeyear{iosif2022large}) investigated security vulnerabilities in AWS repositories where the Terraform component is major. 
However, besides only focusing on vulnerabilities, in real-world projects, IaC components tend to co-exist with other types of files and represent only a small percentage of the full project \citep{jiang}. 
To investigate the adoption of best practices related to security policies by Infrastructure as Code (IaC), we propose, categorize and check the implementation of security policies in repositories adopting IaC. 

For that, we identify, select and propose a categorization of recognized security policies, then examine the degree to which each type is observed in real-world deployments.
Next, based on the popularity of IaC environment on GitHub, we perform an empirical study evaluating the implementation of the previous policies on GitHub repositories. 
For that, we mine open-source repositories based on three different cloud providers (AWS, Azure, and Google Cloud), and then we use checkov, a static analysis tool, to scan each project, looking for the presence or absence of the security policies.
Finally, we collect GitHub measures to check whether these measures are correlated with projects adopting these security policies (number of stars, forks, and contributors).

To summarize, we investigate the following research questions in this study:
\begin{enumerate}
    \item \emph{\textbf{RQ1:} Which security guidelines have matching Terraform implementation policies, and how can they be categorized?}\\
    The objective is to select and categorize Terraform security practices that are recognized by the industry. 
    Studying the adoption of security policies alone might lead to restricted results. 
    This way, we decide to perform an analysis based on categories, as it allows generating knowledge on policy patterns, and practitioners' awareness and enables us to compare results. 
    As a result, we report a catalog of eight categories associated with security policies derived from previous work, as also new ones identified by us. 
    
    \item \emph{\textbf{RQ2:} How are common security best practices being adopted in Terraform files?}
    In this RQ, we focus on practices that tend to be well adopted in projects with Terraform components to spotlight patterns that foster implementation. 
    Additionally, we also intend to investigate configuration flaws through neglected practices to suggest improvements in those areas from providers and practitioners. 
    We observe, based on the previous categorization, that the category \textit{Access policy} emerges as the most widely adopted in all evaluated cloud providers.
    For AWS and Azure providers, we observe that \textit{Hard-coded secrets} policies are well implemented, highlighting the impact of cloud services design on security practice through default configurations. 
    Since we consider in our evaluation different cloud providers, we aim to compare the findings obtained through the previous questions between cloud providers in order to generalize and validate some findings, as well as highlight reasons for observation from the characteristic differences between providers.
    
    \item \emph{\textbf{RQ3:} Is there a correlation between the popularity of a GitHub repository and the adoption of security best practices in its Terraform component?}\\
    In this research question, we collect GitHub repositories metadata to explore potential correlations between popularity metrics and the adoption of infrastructure security policies. 
    As repository platforms are often the source for code collaboration and code reuse, this type of knowledge can be useful and insightful to practitioners.
    In the end, we observe a positive, strong correlation between a repository number of stars and adopting practices in its cloud infrastructure.
\end{enumerate}


Based on our findings, we provide general guidelines for cloud practitioners to limit infrastructure vulnerability and discuss further aspects associated with policies that have yet to be extensively embraced within the industry.
As contributions from our work, we may highlight the categorization of security policies related to IaC based on standard industry-recognized patterns, which could be used for future studies and also be extended. 
Next, we perform an investigation exploring the adoption of these policies in Terraform files based on three different cloud providers.
Additionally, we investigate the correlation between GitHub measures and best practice policy adoption.
Finally, we provide our datasets and our method to evaluate the implementation of security policies by checkov online, in order to support further studies as also replications of our current work.

\vspace{5mm}
\textbf{Data Availability Statement.} The datasets generated during the current study are available in our online Appendix \cite{thesis_replication_package}.

\vspace{5mm}
\textbf{Paper Organization.}
The rest of the paper is organized as follows.
In Section~\ref{section2}, we motivate our study and present our research questions. 
Section~\ref{section3} explains our study setup, which is responsible for defining the categorization process, its validation, the GitHub mining step, and the checking of security policies using checkov.
In Section~\ref{section4}, we present the results, which are further discussed in Section~\ref{section5}. 
The threats to the validity of our work are discussed in Section~\ref{section6}, while Section~\ref{section7} discusses related work. 
Finally, in Section~\ref{conclusion}, we present our conclusions.

\section{Motivation and Background}
\label{section2}
This section presents background information and motivates our work with context and real-world examples. 
We will also present the main data protection regulations with which most companies aim to comply to. 
Then, cloud-related security concerns will be tackled with common security best practices and an introduction of the most used security industry standards. 
Finally, we will talk about Infrastructure as Code, providing more details about Terraform, which will be the IaC tool we focus on in this paper.

\subsection{Security}

When working with cloud computing environments, security is one of the key challenges to address as more and more sensitive data and critical computation are being moved to the cloud. 
Therefore, it is essential that those systems are protected from malicious attacks and misuse. 
In order to address those problems, data protection regulations have been introduced, and security standards implemented to guide organizations and practitioners on how to build secure systems.

\subsubsection{Data Protection Regulations}

Data protection regulations have existed for several decades. 
However, since the 2010\textsuperscript{th}, those regulations have become more strict regarding the rights citizens have over their data. 
In 2016, the EU (European Union) introduced the GDPR (General Data Protection Regulation) \citep{european_commission_regulation_2016}. 
Implemented since 2018, the regulation applies to any entity using data from a person based in the EU and stipulates that the data collected can be easily accessible.
Besides, each citizen can have access to their personal data and information about how it is being processed. 
The regulation also specifies the right to rectify and erase personal data and to object to the processing of someone's data, e.g., each organization requires the consent of the data subject to store and process its data. 
Finally, GDPR \citep{european_commission_regulation_2016} enforces the use of pseudonymization when storing and processing personal data and the creation of records for every activity processed.

While GDPR protects European citizens, it became a template for several other data protection regulations worldwide, such as CCPA (California Consumer Privacy Act) \citep{ccpa}, Canada Privacy Act, General Personal Data Protection Law in Brazil, and AAPI (Act on the Protection of Personal Information) in Japan.
With the fast-growing number of users under those regulations, organizations are obliged to comply to avoid withdrawing from those territories and their citizens as potential customers, which requires secure and compliant computing in every industry.

\subsubsection{Cloud Security Industry Standards}
In addition to these laws, there are also various industry standards and best practices that organizations can follow to enhance the privacy of their cloud systems. 
It should also be noted that since the introduction of regulations, security standards have also been extended to include regulation compliance. 
In this study, we will focus on the two standards considered the most useful among cloud practitioners \citep{stultienscompliant}.

To help practitioners comply with cloud infrastructure security and privacy regulations, providers and independent organizations have created sets of best practices and security benchmarks. 
Some form the basis of security certifications, of both cloud deployment audits and practitioners' knowledge. 
They help build customer trust and have become industry standards for secure cloud infrastructure \citep{auditing, sec_engin}. 
The two main standards for AWS are the Center for Internet Security (CIS) AWS Foundations Benchmark \citep{cis} and the AWS Foundational Security Best Practices standard \citep{aws_fundamentals}. According to Stultiens \citep{stultienscompliant}, those two standards are widely respected and adopted. 
The CIS also developed security best practices standards on Azure \citep{cis_azure} and Google Cloud \citep{cis_google}.

\subsubsection{Security Best Practices}

In this subsection, we will introduce the most common security practices that are leveraged in cloud deployments.

\subsubsection*{Enforcing encryption}
Encryption is a process to conceal data in order to restrict which party can access and understand the data. 
With cryptography protocols, it converts plain information into incomprehensible ciphertext. 
Most of the time, cryptographic keys are used to encrypt the data and are needed to recover the original information. 
Encrypting data can help ensure the confidentiality, security, and integrity of the data by making the data unintelligible to those who do not have access to the encryption key. 
We distinguish symmetric encryption using one key to encrypt and decrypt data from asymmetric encryption using a pair of keys (generally a public key and a private key). 
Symmetric encryption is fast, efficient, and often used to encrypt data at rest (i.e., data stored on a device). 
In transit, a hybrid approach is often employed to securely establish an encrypted connection, where asymmetric encryption is used to exchange a shared symmetric encryption key. 
TLS/SSL is the main protocol used to securely exchange data in a network and is based on asymmetric encryption.

\subsubsection*{Preventing admin privileges by default}
In order to avoid granting all privileges to every user of a system, the introduction of permission roles is primary. 
Admin (short for administrator) is the role possessing all privileges on a system. 
Therefore, having access to administrator privileges by default can represent serious vulnerabilities, as any malicious user with access to the role has the power to make major changes. 
Preventing that admin privileges from being granted by default avoids this common type of vulnerability. 
As a rule of thumb, we often call the least privilege principle the practice of only allowing the strictly required privilege to perform one desired purpose.

\subsubsection*{Restricting access policies}
Computing systems and infrastructures are often composed of several sub-systems interacting with one another.
It is a good practice to restrict the access a sub-system has to other subsystems and only allow strict minimum communications. 
Most of the time, we use policies to describe the access each sub-system has to the other sub-systems. 
The least privilege principle also applies to access policies by only granting the strictly required access authorization to perform the task. 
Over-permissive access policies can lead to unexpected access to subsystems and perform malicious actions.

\subsubsection*{Use of logs and monitoring tools}
Monitoring different metrics, such as resource usage and network activity on a system, can help identify patterns and detect anomalies. 
Likewise, logging data can help identify the source of issues or trace back the events of an attack. 
While not a preventive security defense measure, anomaly detection in logs and monitoring metrics can quickly detect security breaches and proceed with more strict security measures to limit the scope of the attack. 
Afterwise, the logs can be very helpful to improve the current security setup.

\subsubsection*{Using last software versions}
Cybersecurity is an endless issue between attackers and defenders. 
Vulnerabilities in software are discovered and fixed every day through security updates. 
By using previous and deprecated software in its system, the risk of attacks from unpatched software vulnerabilities rises dramatically. 
Staying up to date by enabling automatic minor updates limits this issue by automatically applying security patches.

\subsubsection*{Non-disclosure of hard-coded secret}
With the predominant use of VCS (Version Control System) and code hosting platforms like GitHub used to collaborate on software projects, it is very common to upload part of the source code publicly. 
However, the source code may contain secrets such as passwords, private encryption keys, or access tokens to proceed with its task. 
By publicly publishing source code to the internet, these secrets hard-coded in the source files can be extracted and maliciously reused. 
Therefore, before uploading new source code publicly, one must ensure its code does not contain any secrets granting extended access to malicious attackers.

\subsection{Infrastructure as Code}
IaC is the concept of defining computing infrastructure requirements with source files in the form of code. 
Then, the tool understands the defined infrastructure and automatically deploys the environment in the cloud and/or on-premises environment. 
This process permits the infrastructure definition also goes through the DevSecOps lifecycle, enabling compliance and security checks before deployment.
As popular IaC tools, we may cite Terraform, Chef, Puppet, Ansible, and CloudFormation.

\subsubsection{IaC Tool Types}
IaC tools can address different types of needs and layers in an infrastructure.
On the one hand, IaC tools can create, modify and destroy infrastructure resources like computation instances, storage, and networking components. 
Those tools are Infrastructure Management tools. 
Other tools can be used to deploy and update the application running on the infrastructure \citep{morris2020infrastructure}. 
Most popular frameworks can be used to not only deploy the infrastructure but also the application. 
It is possible broad tools use more specific components to take care of the infrastructure management step, like Ansible can use Terraform for the infrastructure needs \citep{nayak_2019}. 
Our study focuses on the infrastructure management stage of IaC, which directly interacts with cloud providers to define and provision cloud infrastructures. 
In previous work, Guerriero et al. (\citeyear{guerriero2019adoption}) studied the adoption of IaC tools in the industry. 
Table \ref{tab:guerriero_table} shows the adoption results they obtained, in which we see Terraform as being the most popular IaC tool focusing on infrastructure management. 


\begin{table}[htbp]
    \centering
        \caption{Adoption of IaC tools by Guerriero et al. (\citeyear{guerriero2019adoption})}
    \begin{tabular}{c|c|c}
Tool & Usage & Type of IaC\\
\hline
Docker & 59.0\% & Container and templating\\
Ansible & 52.2\% & Configuration management\\
Kubernetes & 40.9\% & Container and templating\\
Chef & 36.3\% & Configuration management\\
Terraform & 34.1\% & Infrastructure management\\
Puppet & 29.5\% & Configuration management\\
\rule{0pt}{0pt}
[...] & [...] & [...] \\
CloudFormation & 20.0\% & Infrastructure management\\
Shell scripts & 09.0\% & Multipurpose\\
Azure DevOps & 02.3\% & Infrastructure management\\
    \end{tabular}
    \label{tab:guerriero_table}
\end{table}

\subsubsection{Terraform}
Terraform \citep{terraform} is a leading infrastructure management tool with extensive compatibility across providers and integration with other tools, like Ansible. 
It is an open-source project led by HashiCorp, released in 2014. 
The tool supports managing infrastructures from public cloud providers such as AWS, Azure, and GCP, as well as private cloud frameworks like OpenStack. 
Terraform files' main purpose is to declare the infrastructure resources. 
Infrastructure provisioning lets users automatically deploy and configure resources like servers, storage, networks, and services directly in the cloud \citep{juve}. 
Resources are declared in blocks, with all the relevant configurations within Terraform files written in HCL (HashiCorp Configuration Language). 
For example, a block declares a computing instance with the required computing power, the size of the allocated, and the image to deploy on it. 
Modules can be used to ease the definition of the infrastructure by reusing code, like programming libraries. 
Once the infrastructure is declared, Terraform converts the code into API calls to deploy the resources. 
The tool acts as an interface between the cloud practitioners and the cloud provider API to ease the deployment.

\begin{figure}[htbp]
\centerline{\includegraphics[width=\linewidth]{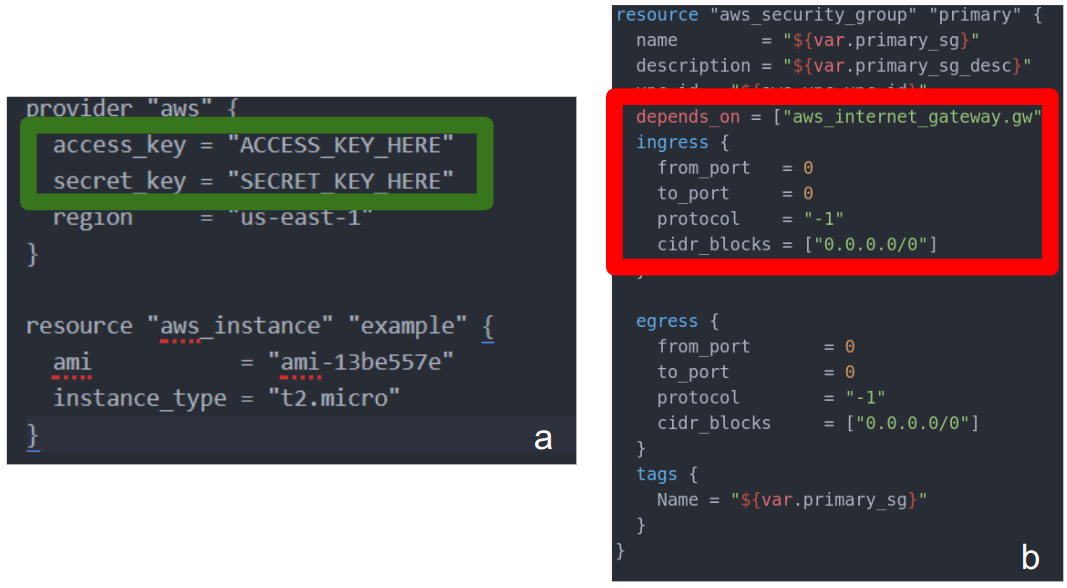}}

\caption{Terraform Definition Examples of the following security policies:\\
a. Passing Non-disclosure of secrets: \\{\color[HTML]{229954}Ensure no hard-coded AWS access key and secret key exists in provider}\\
b. Failing Restricting access policies: \\{\color{red}Ensure no security groups allow ingress from 0.0.0.0:0 to any port}
}
\label{fig:terraform_examples}
\end{figure}

Figure \ref{fig:terraform_examples} presents two examples of Terraform definitions and the related security policies that can be looked for. 
In Figure \ref{fig:terraform_examples}.a, no hard-coded secrets are contained in the provider definition, which indicates a good adoption of the \emph{Non-disclosure of secrets} practice. 
However, in Figure \ref{fig:terraform_examples}.b, the security group configuration allows ingress traffic from any source (0.0.0.0:0 IP address binding), which indicates that the security practice \emph{Restricting access policies} have not been implemented. 
Those two examples show, in practice, the type of configuration we will be looking for when trying to measure the adoption of security practices by practitioners in Terraform files.

\subsection{Study Motivation}

As data regulations have been introduced to protect the rights of consumers' data, industry compliance standards have been designed to guide cloud practitioners on how to build secure infrastructures. 
However, we may wonder whether practitioners actually implement those guidelines when building real-world applications.
Recent massive data breaches due to misconfigured cloud infrastructures suggest that compliance to those standards is not perfectly implemented everywhere \citep{bluebleed}. 
Therefore, our study aims to gain knowledge on the adoption of security standards practices by practitioners on real-world projects deployed on popular cloud services providers. 
The goal of this paper is to quantitatively measure the adoption of security practices by practitioners through Static Code Analysis of Terraform components and extract meaningful knowledge on practice adoption across different cloud providers. 
While few previous studies have been processed in this direction (see Related Work in Section \ref{section7}), none focus on adopted as well as neglected practices of real-world projects deployed on either one of three different cloud providers (AWS, Azure, and GCP).


\section{Study Setup}
\label{section3}
To address the previous research questions, in this section, we present the methodology adopted in our study. 
Overall, our methodology comprises two main steps (Figure~\ref{study_method}). 
First, we select and categorize Terraform security practices promulgated in the industry for popular cloud providers, such as AWS, Azure, and Google Cloud.
Second, to address RQ2, we select open-source repositories hosted on GitHub in order to assess the adoption of the previously categorized practices.
Finally, to answer RQ3, we collect additional metrics associated with our sample of repositories in order to investigate factors that might be correlated with adopting these best practices. 
The scripts used to automate these steps are available in our online Appendix~\citep{thesis_replication_package}, supporting further replications and new related studies.



\subsection{Policies Categorization}
\label{sec:methodology-rq1}

\begin{figure}[htbp]
\centerline{\includegraphics[width=\linewidth]{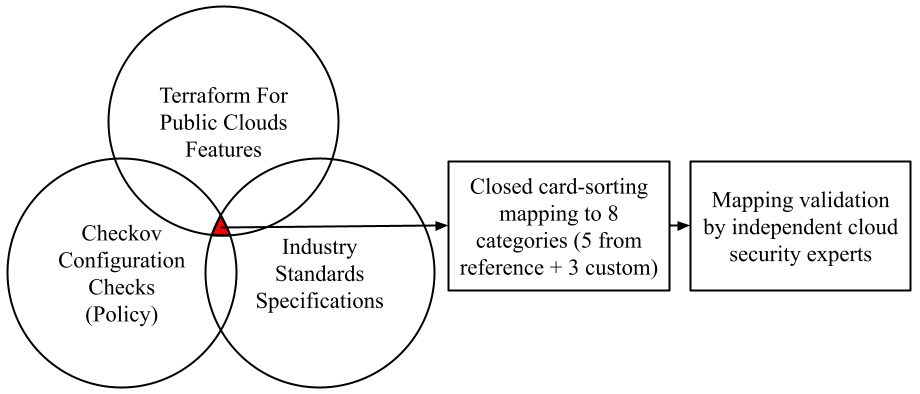}}
\caption{RQ1 Policies Selection And Categorization Process}
\label{rq1_method}
\end{figure}

Security best practices are general concepts and techniques whose implementation differs from cloud providers, types of infrastructures, and configurations, such as IaC tools. 
Therefore, security best practices can be represented as a set of policies, which are actual code snippets implementing security guidelines. 
Static code analysis (SCA) can be processed on Terraform files to check the implementation of specific policies related to best practice types (or categories). 
However, not all policies built-in SCA tools are relevant. 
This way, to address our first research question (RQ1), we carefully select and compile a list of industry-recognized security policies, mapping them to their related security best practices. 
Once these policies are selected, we adopt a closed card sorting method to categorize the selected policies based on new categorizations and previously proposed categories \citep{rahman}. 
After that, we cross-validate the mapping (policy categorization) with independent cloud experts. 
The method is synthesized and visualized in Figure \ref{rq1_method}.

\subsubsection{Standard Guidelines: Industry-Recognized Security Policies}

For each cloud provider evaluated in this study, we use a systematic approach to select built-in policies from a list of industry-recognized security policies with known Terraform implementation signatures. 
For the AWS cloud provider, we observe that the CIS Amazon Web Services Foundations~\citep{cis}\footnote{Version v1.4.0} and the AWS Foundational Security Best Practices\citep{aws_fundamentals}\footnote{Version v1.0.0} are the two most widely recognized and deployed industry-standard frameworks~\citep{stultienscompliant}. 
For the Azure, the Center for Internet Security (CIS), which established the CIS Amazon Web Services Foundations, also defined an Azure equivalent, the CIS Microsoft Azure Foundations Benchmark~\citep{cis_azure}.\footnote{Version v2.0.0}
So we manually go through these 298 policies to select those that could be mapped to a specification of one of the two standards, selecting 121 policies. 
For the Azure provider, 247 security configurations are tagged to Azure deployments on Terraform. 
Finally, for the Google Cloud provider, we consider the CIS Google Cloud Platform Foundation Benchmark~\citep{cis_google}. 

\subsubsection{Validation by Independent Experts}
Since our mapping relies on the subjective judgment of one researcher, we mitigate potential bias by inviting independent Security Consultants for each cloud provider.
We generally look for experts with a solid background in cloud configuration and security policies that are not authors of this study.
Once the consultants were selected, instructions were given by video conference, and possible questions were also addressed. 
Next, we provided the experts with a description of each policy and the category previously mapped in the form of a spreadsheet. 
Additionally, they had access to the full documentation of the policies, with their code implementations as provided by Checkov. 
For each policy (each row of the spreadsheet), the experts were asked to agree or disagree with its mapping by checking an empty checkbox. 
After that, their individual files were evaluated, and a Cohen's Kappa score was measured to assess their conformity.

\subsection{Selecting Open-Source Repositories based on Cloud Providers}
\begin{figure}[htb]
\centerline{\includegraphics[width=.7\linewidth]{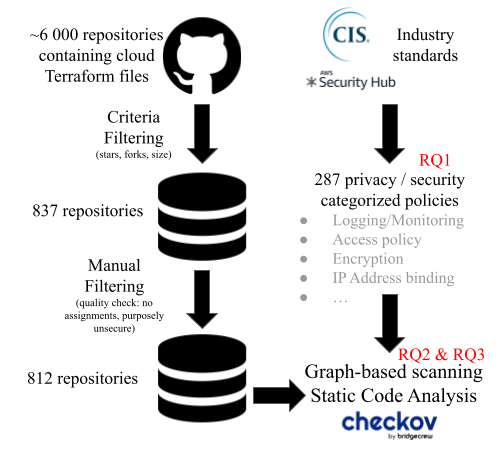}}
\caption{Research Process Overview}
\label{study_method}
\end{figure}

In order to assess the adoption of IaC security best practices, we evaluate the implementation of the previously selected policies.
For that, first, we systematically select repositories hosted on GitHub based on changes in specific configuration files. 
Next, we filter these projects based on some metrics. 
Based on each cloud provider, this process was repeated, leading to the sample of projects evaluated in this study (Figure \ref{study_method}). 

\subsubsection{Datasets Collection} \label{data_collection}

This section presents the method for collecting our dataset of real-world projects deploying the different cloud provider infrastructures from GitHub. 
Jiang and Adams (\citeyear{jiang}) report that IaC components co-exist with other types of files in open-source projects. 
The median of IaC files turned out to be around 11\% of the total project file number. 
Therefore, the most representative way to collect relevant repositories is by searching for Terraform files corresponding to our chosen cloud provider. 
This helps collect repositories, which the cloud infrastructure is not the main topic addressed in GitHub projects' titles and descriptions. 
The usual GitHub Search API, as used by Iosif et al. (\citeyear{iosif2022large}), only allows searching for keywords in repositories titles, topics, or README files. 
It is possible to filter repositories by programming language (with the snippet ``language:HCL''), but this approach only retrieves repositories where the searched language corresponds to the majority of the project (e.g., $>$ 50\% of the project size). 
Instead, we need to use the more restrictive GitHub \textbf{Code} Search API \citep{github_code_search_api}, which allows us to search for code snippets inside the source files.

\begin{figure}[htb]
\centerline{\includegraphics[width=.7\linewidth]{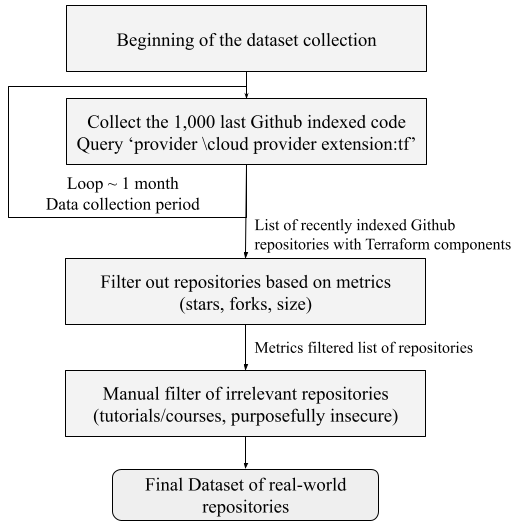}}
\caption{Terraform GitHub Projects Collection Process}
\label{data_collec_process}
\end{figure}

Terraform files are widely used in GitHub repositories. 
For example, as of October 2022, the GitHub research query looking for usage of Terraform files on AWS counts more than 350K code results (\texttt{provider aws extension:tf}). 
However, collecting the corresponding Terraform files and repositories is hard for several reasons. 
First, cloud computing advances rapidly, and snapshots of large GitHub databases such as the GHTorrent \citep{ghtorrent} contains many outdated projects that may not reflect current practices. 
Second, no standard tag or label can be used to decide whether a repository contains an AWS infrastructure deployment configuration. 
Third, using repository search with keywords like `Terraform AWS' will either not return the intended GitHub repositories (e.g., projects deployed to the cloud with an IaC component) or will return false positives (e.g., tutorials/course repositories). 
And finally, the GitHub API only allows access to the first 1,000 results (sorted by relevance or indexation date). 
To manage these constraints, we adopted the process depicted in Figure \ref{data_collec_process}.

The standard file extension for Terraform file is `.tf'. 
This way, we expect most of the targeted files to follow this guideline. 
Moreover, the selected cloud provider has to be introduced in one file with the code snippet ‘\texttt{provider “cloud\_provider”}’. 
As such, for each cloud provider, we query the GitHub Code Search API and successfully collect the last 1,000 indexed corresponding Terraform files. 
The results are sorted by \emph{Last indexed files} to avoid bias from GitHub \emph{Best match} algorithm. 
Since 1,000 Terraform indexed files at a certain point in time become no longer representative, we ran the same query and collected the repositories periodically each day for specific consecutive days.
Eventually, duplicated repositories were removed, leaving only distinct repositories.

For the AWS provider, we run the associated query for three weeks (from September 12th to October 5th, 2022). As a result, 3,245 distinct GitHub repositories with at least one AWS Terraform file were selected.
For the Azure provider, the data collection period also lasted one month (from January 20\textsuperscript{th} to February 20\textsuperscript{th}, 2023), reporting 1,308 repositories.
Finally, for the Google Cloud provider, the data collection period lasted one month from March 20\textsuperscript{th} to April 20\textsuperscript{th}, 2023. In total, 1518 repositories were selected.

\subsubsection{Repository Filtering}
\label{sec:repository-filtering}
With the initial collected projects, we next filter the repositories, selecting those relevant to our study. 
For that, we adopt the same filter metrics adopted by previous studies \citep{das,gonzalez}, that examined GitHub repositories for Blockchain and Machine Learning projects, respectively. 
This way, we collect the next metrics \citep{gousios}:
\begin{itemize}
    \item Activity: the last project indexation must be later than September, $1^{st}$ 2022. 
    \item Size: a repository must have a non-null size (e.g., $>$ 0 KB). 
    \item Originality: a repository must not be a fork of another project. 
    \item Popularity: a repository must have at least $\geq$ 2 stars. 
    \item Data Availability: the repository must be publicly accessible via the GitHub API.
    \item Content: the project must not be a course assignment, tutorial, or purposefully insecure project.
\end{itemize}

The first criterion is inherent to our sample collection method, while the sixth criterion requires manual filtering. 
For the AWS sample, after applying the first five criteria, 413 were remaining. 
Next, after manually analyzing each project, 12 of which we removed, leading to the final sample of 401 relevant repositories. 
More specifically, we remove two (2) purposefully insecure projects, six (6) workshops/assignments, and four (4) course materials/tutorials. 
For the Azure sample, after applying the first five criteria, 143 projects were returned. 
Next, after the manual analysis, 6 projects were removed, leading to the final sample of 137 Azure projects. 
The discarded repositories refer to workshops/assignments (3) and course materials/tutorials (3).
Finally, for the Google Cloud sample, 281 repositories passed the first five criteria.
After the manual analysis, 274 repositories compose the final sample. 
Regarding the manual analysis, the removed repositories refer to workshops/assignments (4), course materials/tutorials (2), and purposefully insecure projects (1).

We remark that the final Azure and Google Cloud samples are smaller than the AWS one. 
In March 2023, 417K Terraform files with ``\texttt{provider aws}'' and 101K with ``\texttt{provider azurerm}'' were indexed on GitHub, while the GitHub query ``\texttt{provider google}'' only shows 80K indexed files.
The Azure sample size difference seems to fit the overall GitHub population from which they have been collected. 
However, the number of repositories in the final GCP dataset does not match the GitHub population relative to the two other datasets.
This can be explained by the popularity metrics that tend to be statistically higher on the repositories with GCP Terraform components, leading to more repositories passing the popularity filter.

\subsubsection{General Metrics Collection}

After establishing the sample of our empirical study, we further explore metrics associated with each project. 
This way, using the GitHub API, we explore the number of stars, forks, and contributors to a project.
Based on the results of RQ2, we aim to assess how these metrics correlate with the adoption or not of the investigated secure policies. 
For that, we check the correlation between each metric and the rates of passing and failing associated with the secure policies using the Spearman correlation test.
Table \ref{table:datasets_metrics_comparison} presents metrics such as the mean and median numbers of stars, forks, and contributors of the final dataset repositories.

\begin{table}[htb]
\begin{center}
\caption{Final datasets metrics comparison}
\begin{tabular}{c|c|c|c}
\textbf{Metric} & \textbf{AWS} & \textbf{Azure} & \textbf{GCP} \\
\hline
Number of GitHub projects & 401 & 137 & 274\\
Mean number of stars & 174.76 & 50.93 & 281.31\\
Median number of stars & 7.0 & 6.5 & 9.5\\
Mean number of forks & 43.26 & 56.07 & 95.37\\
Median number of forks & 4.0 & 3.0 & 5.0\\
Mean number of contributors & 7.60 & 9.30 & 9.89\\
Median number of contributors & 3.0 & 4.0 & 4.0\\
\end{tabular}
\label{table:datasets_metrics_comparison}
\end{center}
\end{table}

\section{Results}
\label{section4}
In this section, we report the results addressing our previous research questions. 
First, we select widely recognized Terraform security practices promulgated in the industry for popular cloud providers, such as AWS, Azure, and Google Cloud. Next, we categorize these practices into eight categories, answering RQ1.
Second, to address RQ2, we assess the adoption of the previously categorized practices by selecting 812 open-source projects hosted on GitHub. 
Then, we scan each project’s configuration files through static analysis performed with the Checkov tool, looking for policy implementation and assessing the rates of passing and failing policy checks.
Finally, to answer RQ3, we investigate factors that might be correlated with adopting these best practices. 
For that, we collect some additional metrics and then check whether these metrics correlate with the rate of passing/failing policy checks. 



\subsection{\textbf{RQ1: Which security guidelines have matching Terraform implementation policies, and how can they be categorized?}}
\label{sec:rq1}

In order to quantitatively reason about adopting security best practices in Terraform files, we first have to define (i) the list of target configurations and (ii) the approach to check the occurrence of these configurations.
This way, we check on industry security best practice standards, identifying security configurations relevant to real-world deployments. 
Each configuration, in the form of policies, is then categorized using a close card-sorting approach to be able to extract knowledge on the best practices \citep{spencer2004card, spencer2009card}. 
Next, we use the SCA tool Checkov built-in policy collection to extract Terraform implementable security configurations. 

\subsubsection*{List of industry-recognized security best practices}

Once we selected the standard guidelines, we explored each official documentation collecting the associated security policies. 
For the AWS cloud provider, Terraform presents 298 AWS configurations.
Then, we need to select those reflecting industry-recognized security and compliance best practices. 
So we manually go through these 298 policies to select those that could be mapped to a specification of one of the two standards, selecting 121 policies. 
For the Azure provider, 247 security configurations are tagged to Azure deployments on Terraform. 
Again, we manually go through the 247 policies to select those that could be mapped to the CIS Azure Benchmark, selecting 84 policies. 
Finally, for the GCP provider, 137 built-in policies are tagged to Terraform deployments of cloud infrastructures on the Google Cloud Platform, selecting 82 policies, which could be mapped to the associated benchmark. 


\subsubsection*{Configurations Implementation Supported by Terraform files}
As previously discussed in Section~\ref{section2}, Checkov comes with over 2500+ built-in policies which check security and standards compliance on several cloud providers, such as AWS, Azure, and Google Cloud. 
Each policy corresponds to a specific configuration snippet in the IaC file, implementing a feature in the provisioned cloud resource. 
Checkov can analyze 11 different infrastructure provisioning tools, including Terraform and Cloudformation. 
This way, we explore Terraform files that define cloud infrastructures for the cloud providers assessed in this study. 
In the scope of this study, we do not consider the addition of any custom policies. 

Among the 2500+ built-in policies, 298, 247, and 137 of them are tagged for AWS, Azure, and Google Cloud, respectively, on Terraform. 
The tool ensures that all 682 policies can be implemented and found in Terraform files. Additionally, other practices might be implemented through Terraform configurations that do not have a related Checkov policy. 
However, since Checkov is an open-source tool, it is expected that new policies will be added over time.

\subsubsection*{Category Mapping}
Mapping each configuration into categories is essential to retrieve broad knowledge from the database scanning analysis. 
To achieve this goal, we first start manually categorizing the AWS policies. 
For that, we use a closed card-sorting method, trying to categorize the policies into seven (7) categories originally proposed by Rahman et. al (\citeyear{rahman}).
During the first mapping round, 72 policies were mapped into five categories: (\textit{Hard-coded secret}, \textit{IP Address binding}, \textit{Admin by default}, \textit{Encryption at rest}, \textit{Encryption in transit}.
The remaining 49 policies were further evaluated in a second round, leading to the addition of two new categories: \textit{Access Policy} and \textit{Logging/Monitoring}.

Since our study focuses on Terraform, new categories are required as they focus on the infrastructure provisioning stage of IaC. 
The \textit{Access Policy} category refers to any misconfiguration of Identity and Access Management roles, as well as Access Control Lists. 
Managing Access policy is an important part of cloud infrastructure security, as it defines which resources can access one another. 
Regarding the \textit{Logging/Monitoring} category, Avila et al. (\citeyear{avila2021}), and Vaarandi and Pihelgas (\citeyear{vaarandi}) show that even if the absence of logs does not represent a direct vulnerability to the infrastructure, logs can be very useful to collect metrics and detect security issues. 
Findings in the adoption of logs in cloud security could prompt interesting insights.
As a result, 45 policies could be mapped into the new categories, leading to four (4) policies without categorization. 
The last 4 policies were further evaluated in a third round. 
They refer to vulnerable resource versions and upgrade configurations that could not be mapped to one of the previous seven categories, resulting in the creation of the category \textit{Outdated Feature}. 
Overall, the total number of interactions in the process is 174 = 121 + 49 + 4.

Next, we follow the same approach to map the policies associated with Azure and Google Cloud providers with the previously described categories.
For the Azure provider, all 84 policies could be mapped to one of the eight categories in one round. 
As all policies were assigned to a valid category, we did not have to look for new ones during the mapping process, validating the generalizability of the previous categories. 
For the Google Cloud provider, the 82 manually selected policies have been mapped to seven (7) of the previous eight (8) categories using a one-round closed card sorting method. 
Indeed, none of the selected policies have been mapped to the category \textit{Hard-coded secrets}. 

\begin{figure}[htbp]
\centering
\includegraphics[width=.5\textwidth]{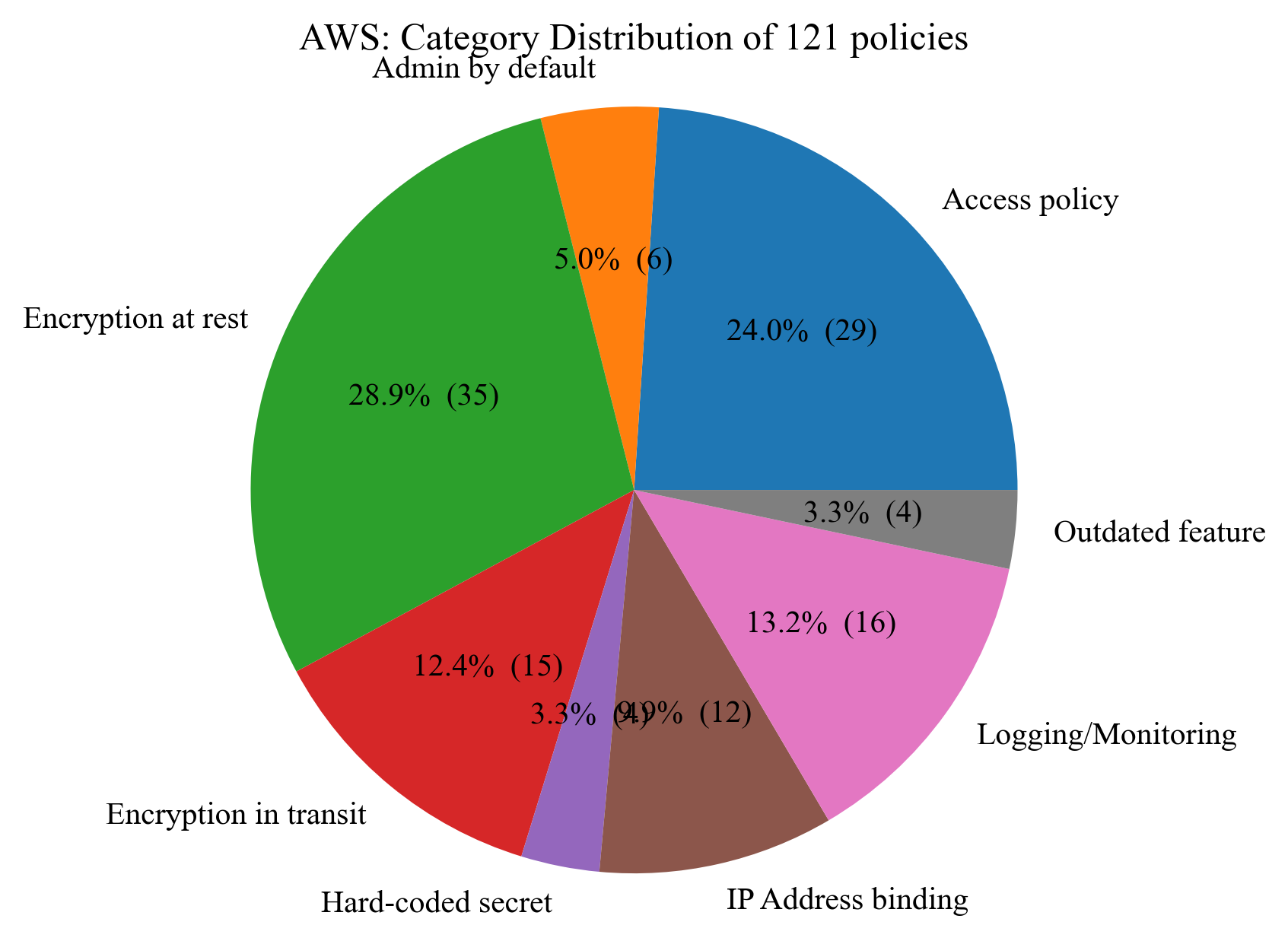}\hfill
\includegraphics[width=.5\textwidth]{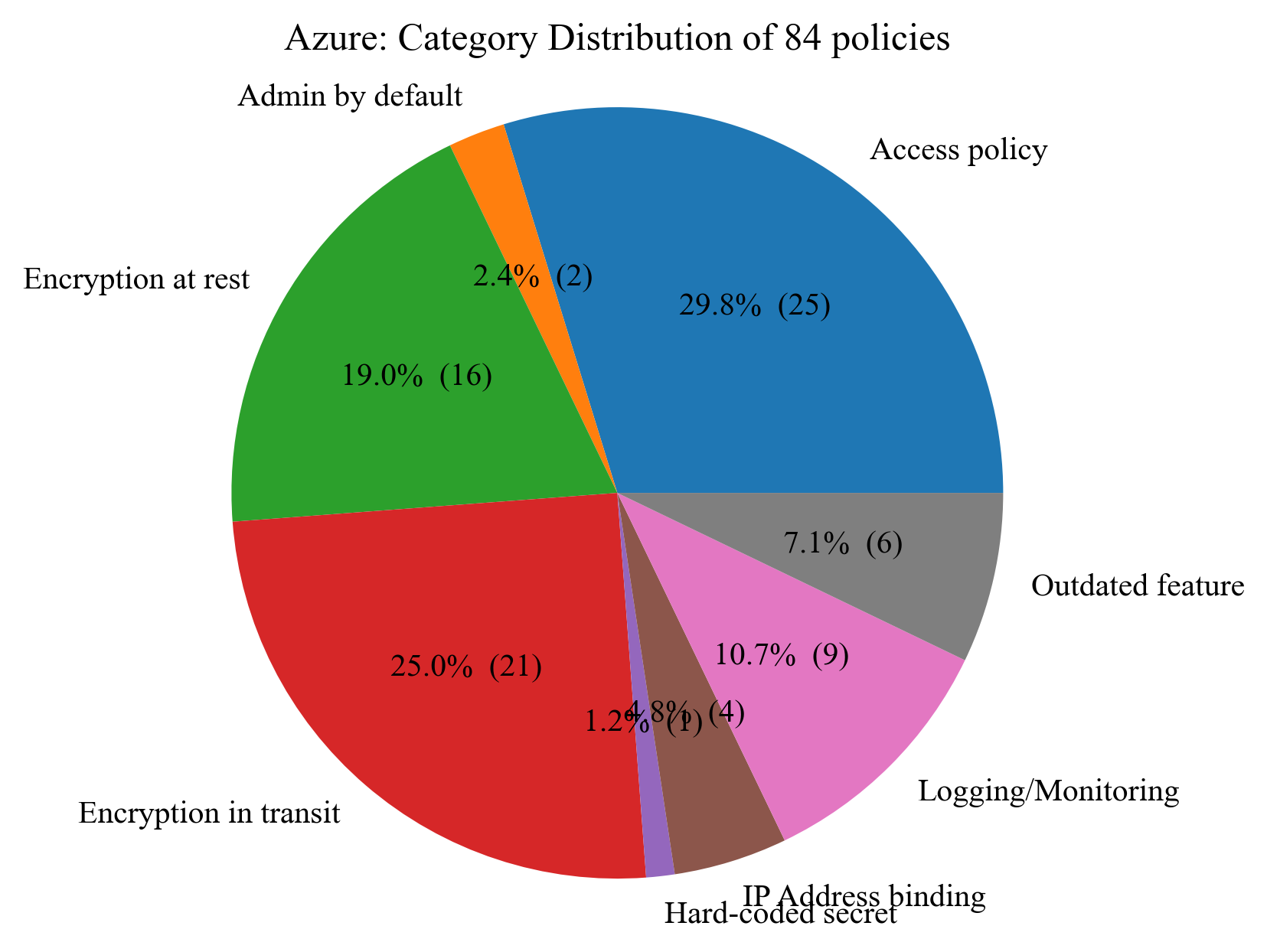}\hfill
\includegraphics[width=.5\textwidth]{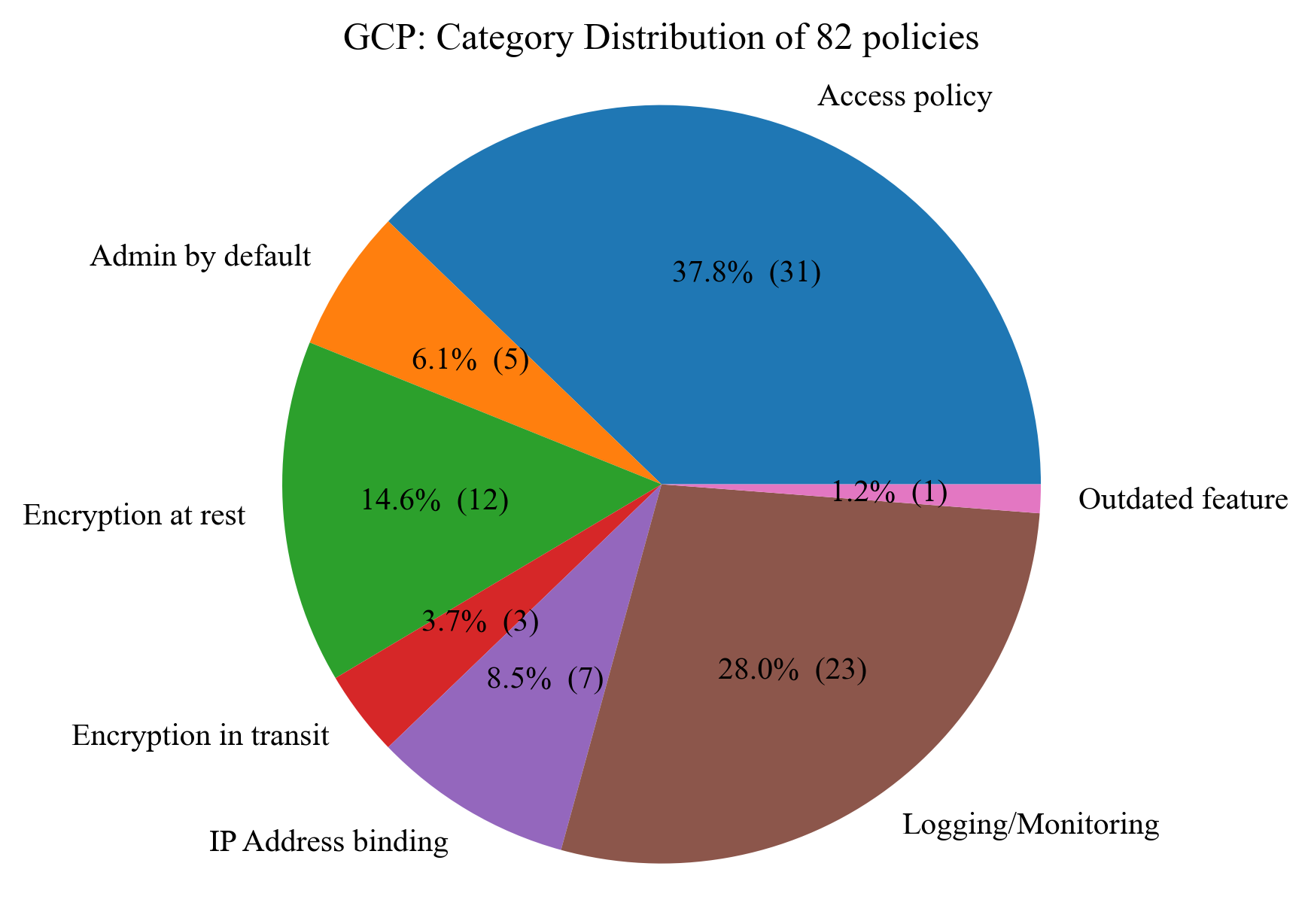}
\caption{AWS, Azure and GCP Taxonomies Distributions}
\label{fig:taxonomies}
\end{figure}

This process was performed by one researcher with strong knowledge of Terraform configuration and cloud providers.
The final distribution of the policies for all providers is presented in Figure \ref{fig:taxonomies}, and the full categorization can be found in our online Appendix~\citep{thesis_replication_package}.

Regarding the validation of the proposed categories, we asked cloud provider experts to check the proposed categories. 
For the AWS provider, two experts were recruited with respectively six and two years of experience, while for the Azure and Google Cloud providers, two other experts were recruited with four years of experience each. 
As a result, we observe a Cohen's Kappa score of $\kappa = 1$, in all validation processes, showing that our categorization represents a valid way to generalize and group security policies based on IaC.

\begin{framed}
\noindent
Our selection and mapping process exhibited relevant configuration snippets in Terraform files for 121, 84, and 82 policies for AWS, Azure, and GCP cloud providers, respectively. 
Those policies are all related to recognized cloud provider security industry standards specifications. 
The policies have been mapped to 8 categories related to common security practices, such as \textit{Hard-coded secret}, \textit{IP Address binding}, \textit{Admin by default}, \textit{Encryption at rest}, \textit{Encryption in transit}, \textit{Access Policy}, \textit{Logging/Monitoring} and \textit{Outdated feature}. 
Figure \ref{fig:taxonomies} shows the category distribution of the policies. 
\end{framed}

\subsection{\textbf{RQ2: How are common security best practices being adopted in Terraform files?}}
\label{sec:rq2}

To answer this research question, we perform an empirical study evaluating the implementation of the previous security policies.
Our sample is composed of 812 GitHub projects divided into the three cloud providers analyzed here, AWS (401), Azure (137), and GPC (274). 
For that, for each repository, we use the static analysis tool Checkov to check the implementation of the selected and categorized policies (see Section \ref{sec:rq1}).
Then, we assess the passing and failing rates for each project.
Before presenting the results for this research question, we discuss next the overall data analyzed during our study.

\subsubsection{Overall Results}

Table \ref{table:overall_results} presents the results for the checking policies performed during our empirical study.
We group the projects based on the associated cloud providers (2-4 columns), and then we report the evaluated metrics (1st column).
Regarding the AWS provider, 59,045 checks are performed exploring the 121 selected policies. 
Among those 59,045 checks, 45,171 have passed while 13,874 have failed, which represents a 76.5\% overall success rate. 
For the GPC provider, 23,127 checks are performed, with a 64\% overall success rate (14,821 passing checks, and 8,308 failed checks).
Finally, for the Azure provider, 5,385 checks are performed based on 84 policies; from these checks, 3,159 presented passing status, while the remaining 2,226 failed, reporting a 58.6\% overall success rate, the smallest success rate among all providers.

Regarding the distribution of checks per policy, overall, we observe the distribution of checks is approximately symmetric based on the \texttt{skew number of checks per policy} (10th row). 
However, for the AWS (0.40) and Azure (0.44) providers, we observe higher values when compared to GPC (0.24). 
This way, we may conclude that for the AWS and Azure providers, the policies do not have the same weight in the results, as some only apply to specific resources rarely deployed. 
Therefore, the category results have been computed from the absolute number of checks per category rather than the average policy results. 
This ensures that each policy contribution to the aggregated results is proportional to its degree of occurrence. 
We also observe that the \texttt{median} and \texttt{average pass rates per repository} (15-16th rows) are close for all cloud providers. 
This suggests that the results are likely to be distributed among the dataset and not biased by a few exceptionally well or bad-performing projects. 
The full results of each policy can be found in our online Appendix \citep{thesis_replication_package}.


\begin{table}[]
\label{table:overall_results}
\caption{Overall Check Results per Provider}
\begin{center}
\begin{tabular}{@{}l|rrr@{}}
\multicolumn{1}{c|}{\multirow{2}{*}{Metrics}} & \multicolumn{3}{c}{Cloud Providers}                                           \\
\multicolumn{1}{c|}{}                         & \multicolumn{1}{c}{AWS} & \multicolumn{1}{c}{Azure} & \multicolumn{1}{c}{GCP} \\ \midrule
Number of GitHub projects                     & 401                     & 137                       & 274                     \\
Number of Policies                            & 121                     & 84                        & 82                      \\
Number of checks                              & 59,045                  & 5,385                     & 23,127                  \\
Number of passed checks                       & 45,171                  & 3,159                     & 14,821                  \\
Number of failed checks                       & 13,874                  & 2,226                     & 8,306                   \\
Median number of checks per policy            & 174.0                   & 20.0                      & 201.0                   \\
Mean number of checks per policy              & 504.66                  & 68.16                     & 282.04                  \\
Skew number of checks per policy              & 0.40                    & 0.44                      & 0.24                    \\
Average Pass/Fail rate per policy             & 0.608                   & 0.5497                    & 0.5627                  \\
Median number of checks per repo              & 56.0                    & 17.5                      & 22.0                    \\
Mean number of checks per repo                & 160.01                  & 52.81                     & 108.66                  \\
Skew number of checks per repo                & 0.33                    & 0.45                      & 0.25                    \\
Median pass rate per repo                     & 71.43                   & 50.0                      & 57.14                   \\
Mean pass rate per repo                       & 69.11                   & 51.90                     & 55.92                  
\end{tabular}
\end{center}
\end{table}

\subsubsection{Commonly Adopted Security Practices}
\label{rq2}

\begin{figure}[htbp]
\centerline{\includegraphics[width=0.9\linewidth]{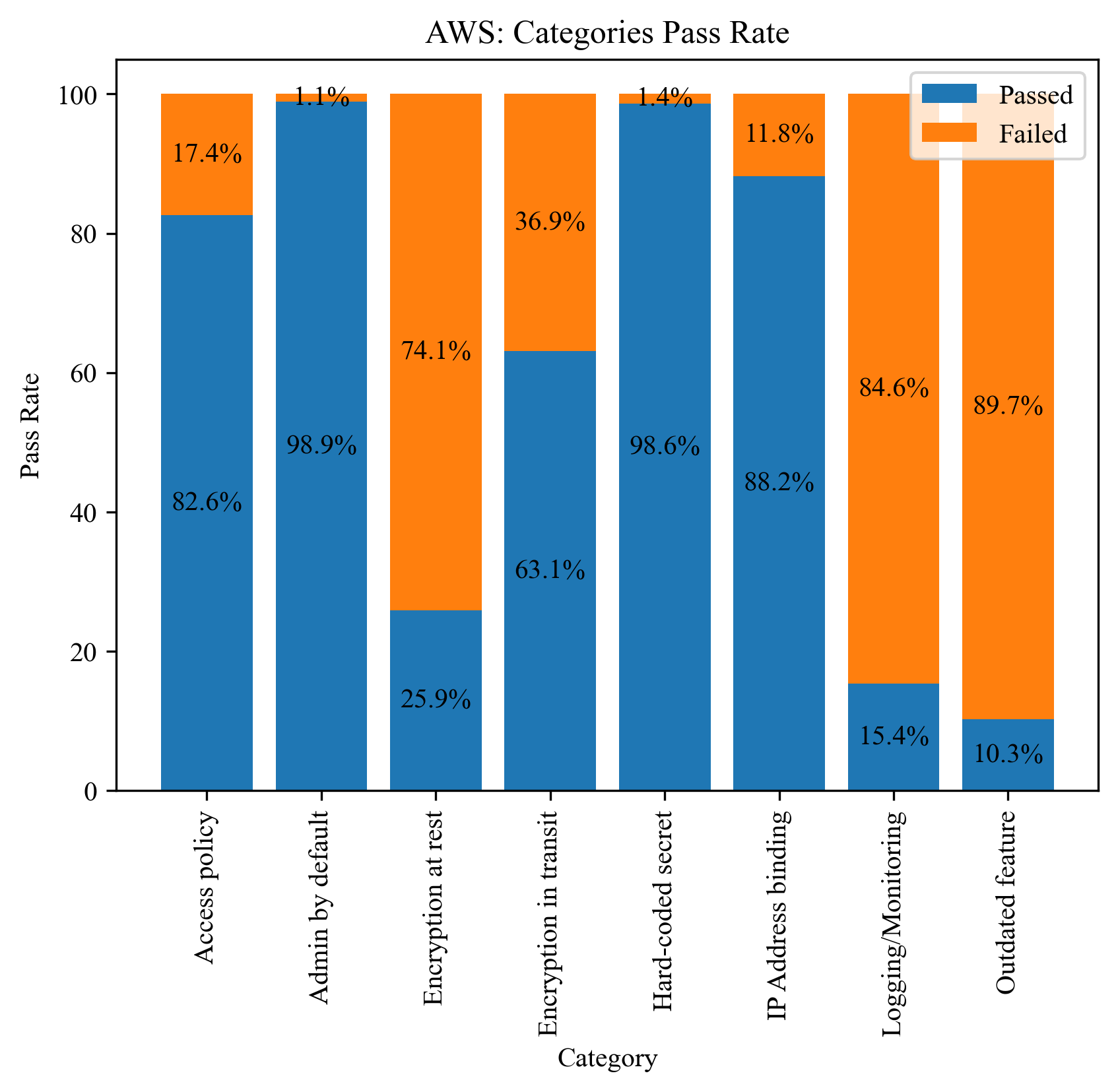}}
\caption{AWS Categories Pass Rates}
\label{categories_results_graph}
\end{figure}

\begin{figure}[htb]
\centerline{\includegraphics[width=.9\linewidth]{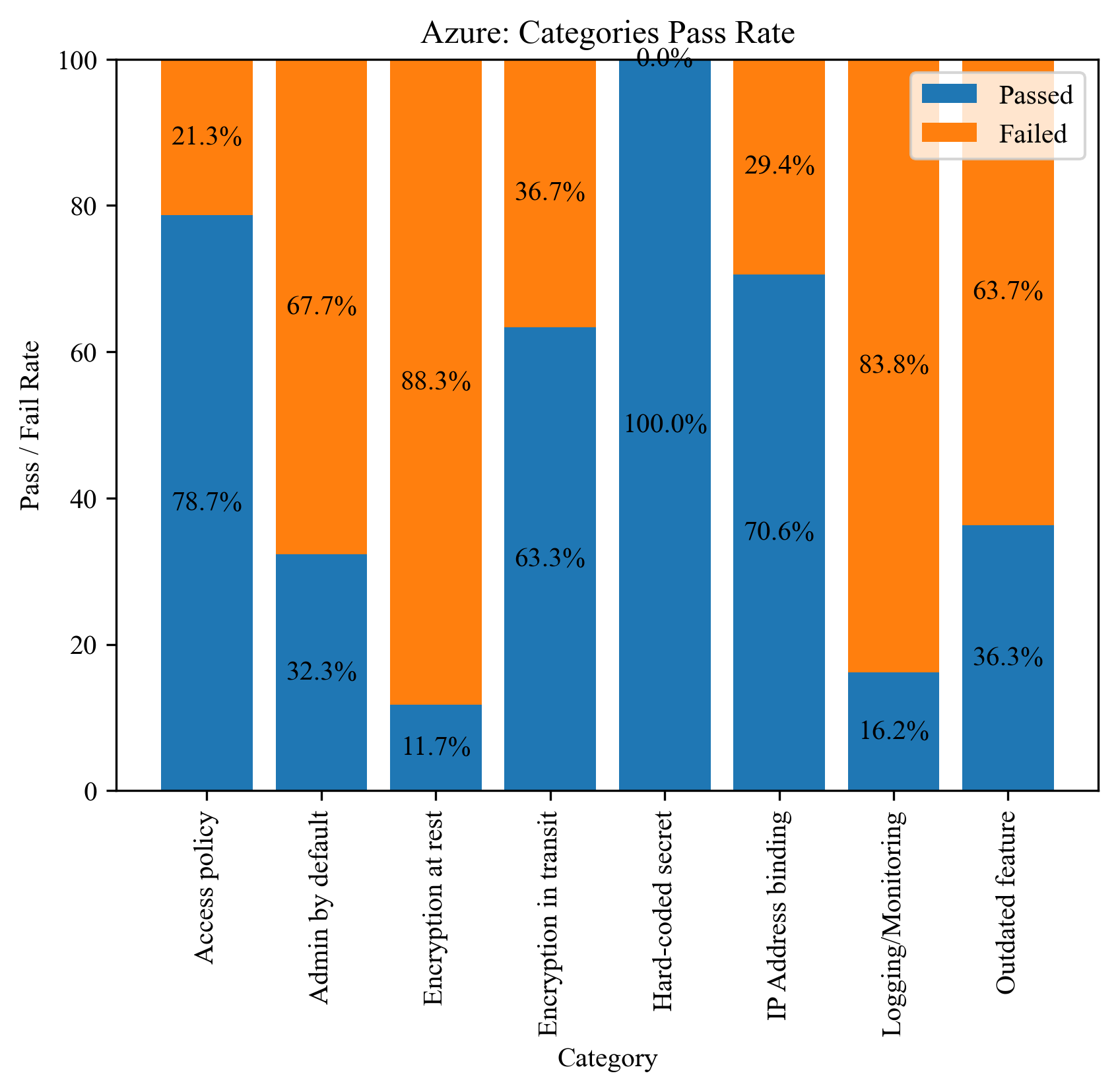}}
\caption{Azure Categories Pass Rates}
\label{azure_cat_results}
\end{figure}

\begin{figure}[htb]
\centerline{\includegraphics[width=.9\linewidth]{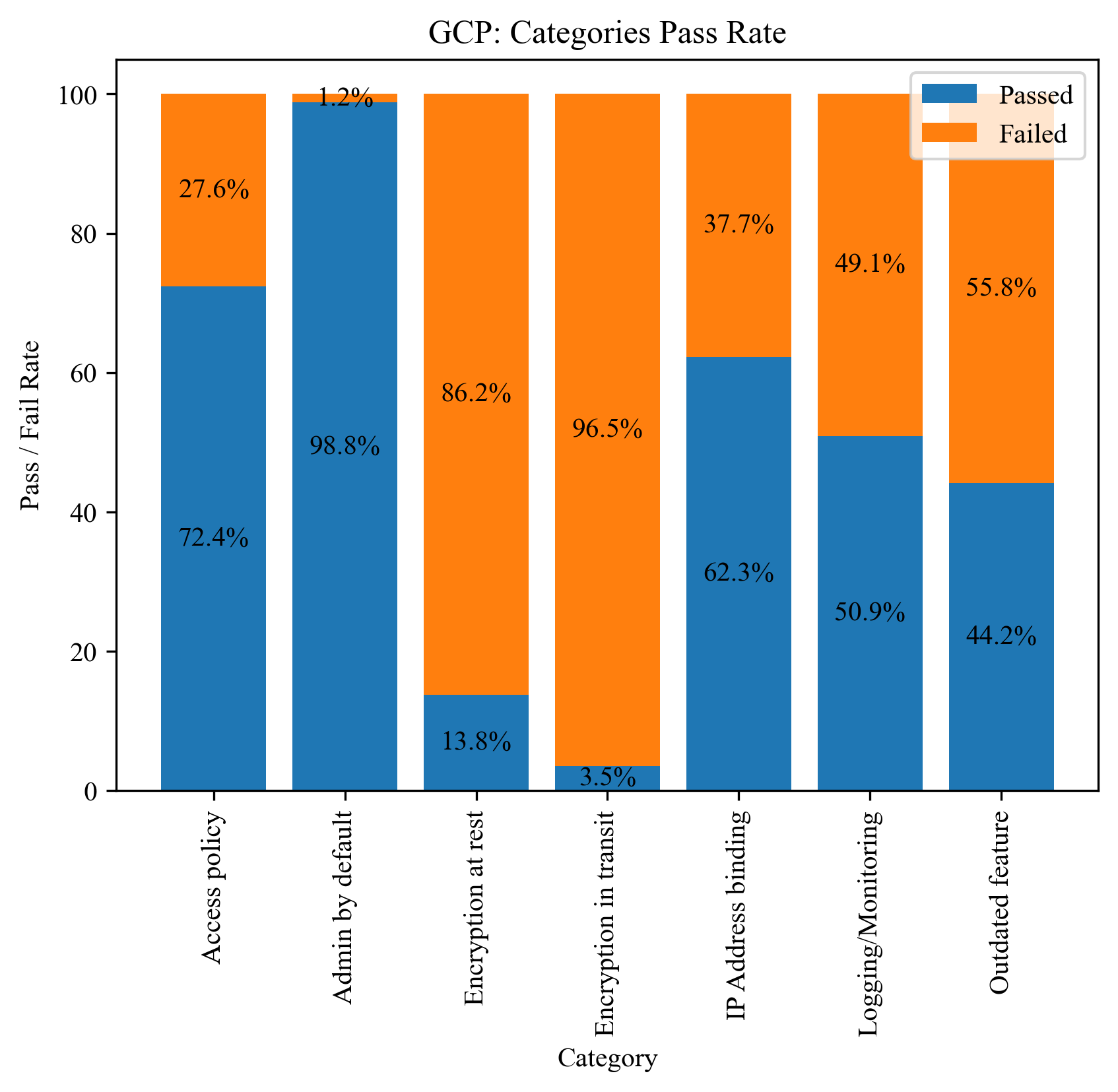}}
\caption{GCP Categories Pass Rates}
\label{google_cat_results}
\end{figure}

In this section, we present the quantitative results obtained by the static code analysis of the cloud providers evaluated in this study.
First, we individually present the results for each provider, and then we compare the findings.
Figures~\ref{categories_results_graph}, \ref{azure_cat_results}, and \ref{google_cat_results} present the overall checking and failing rate for the evaluated policies for AWS, Azure, and GCP providers, respectively.
Tables~\ref{categories_results_table}, \ref{azure_categories_results_table}, and \ref{google_categories_results_table} present more detailed metrics about each category.
Overall, we observed some policy categories perform in all providers, while each provider presented some particularities. 
Next, we discuss individual aspects of each provider and then discuss in detail each of these categories.

For the AWS provider, we can observe that the \textit{Admin by default} and \textit{Hard-coded secret} categories performed very well, with 98.9\% and 98.6\% pass rates, respectively (see Figure~\ref{categories_results_graph}).
Then, we have \textit{IP Address binding} and \textit{Access policy} categories with 88.2\% and 82.6\%, respectively.
The last category with more than 50\% rate is \textit{Encryption in transit} with a 63.1\% pass rate.

For Azure, we observe similar results with AWS. 
The \textit{Hard-coded secret} category does not report any failure (see Figure \ref{azure_cat_results}). 
This category only contains one policy looking for credentials in Virtual Machines data, meaning that no sensitive credentials have been found in the 47 definitions of Azure Virtual Machines with Terraform in our dataset. 
Then, the \textit{Access policy} and \textit{IP Address binding} categories performed similarly, with pass rates of respectively 78.7\% and 70.6\%. 
About \textit{Access policy} SSH, RDP, and HTTP (port 80) accesses, as well as UDP services, are often restricted from the internet (respectively 95.73\%, 89.26\% and 99.77\% pass rate). 
However, disabling public access to Storage accounts is often neglected (with only one implementation among 236 checks). 
The \textit{IP Address binding} only contains 4 policies which show good adoption results, with notably all 47 scanned AKS cluster nodes not having public IP addresses.
Finally, \textit{Encryption in transit} is the last category with a pass rate over 50\% with 63.3\%. 
While \textit{Encryption in transit} is often enabled through HTTPS/TLS, we observe the latest version of those protocols is not always used, especially with Storage Accounts (see our online Appendix for further details~\cite{thesis_replication_package}).

For the GCP provider, just some previous policy categories also present good passing rates here. 
For example, we observe that the \textit{Admin by default} category performs best, with an overall pass rate of 98.8\% (see Figure~\ref{google_cat_results}). 
The second best performing category with 72.4\% pass rate is the \textit{Access policy} category. Table \ref{google_categories_results_table} shows that this category is the one containing the most policies in our GCP policy taxonomy and the one which executed the most checks (9,594). 
Overall, policies restricting access to services to unnecessary parties are well adopted in GCP Terraform files. 
Similarly, the \textit{IP Address binding} category resulted in a 62.3\% overall pass rate. Therefore, we can make the assumption that applying the least privilege principle to Google Cloud services configurations on Terraform is relatively easy for practitioners. 
However, we should bear in mind that such results also highlight that about one-third of the time, those policies are not implemented. 
Potential vulnerabilities might exist in those projects, and improvements can still be made to raise the security of those infrastructures. 
Finally, the \textit{Logging/Monitoring category} shows mixed results with a pass rate of 50.9\%, both on the category pass rate and on the policy level. 
The category contains both the 5 best-performing practices and the 5 worst, which are all related to PostgreSQL databases. 
More investigation on the use of logs in PostgreSQL databases could help better understand those results. 
Besides, we remark that the two most tested policies in this category, VPC Flow and Bucket access logs, are only adopted 15.89\% and 11.77\% of the time, meaning that improvements in this area can be easily achieved.

Next, we discuss these categories in detail.

\begin{landscape}

\begin{table*}[htpb]
\begin{center}
\caption{AWS Policy Categories Check Results}
\resizebox{\linewidth}{!}{
\begin{tabular}{c|c|c|c|c|c|c}
\textbf{Category} & \textbf{Nb of policies} & \textbf{Nb of checks} & \textbf{Nb of pass} & \textbf{Nb of fail} & \textbf{Pass rate} & \textbf{Avg nb of check/policy}\\
\hline
Admin by default & 6 & 9,914 & 9,801 & 113 & 0.9886 & 1,652.3\\
Encryption in transit & 15 & 1,716 & 1,083 & 633 & 0.6311 & 114.4\\
Encryption at rest & 35 & 4,575 & 1,185 & 3,390 & 0.2590 & 130.71\\
Access policy & 29 & 20,510 & 16,993 & 3,577 & 0.8261 & 709.31\\
Logging/Monitoring & 16 & 4,107 & 631 & 3,476 & 0.1536 & 256.69\\
IP Address binding & 12 & 12,766 & 11,263 & 1,503 & 0.8823 & 1,063.8\\
Hard-coded secret & 4 & 4,143 & 4,086 & 57 & 0.9862 & 1,035.75\\
Outdated feature & 4 & 1,254 & 129 & 1,125 & 0.1029 & 313.5
\end{tabular}}
\label{categories_results_table}
\end{center}
\end{table*}

\begin{table}[htb]
\begin{center}
\caption{Azure Policy Categories Check Results}
\resizebox{\linewidth}{!}{
\begin{tabular}{c|c|c|c|c|c|c}
\textbf{Category} & \textbf{Nb of policies} & \textbf{Nb of checks} & \textbf{Nb of pass} & \textbf{Nb of fail} & \textbf{Pass rate} & \textbf{Avg nb of check/policy}\\
\hline
Admin by default & 2 & 96 & 31 & 65 & 0.3229 & 48.0\\
Encryption in transit & 21 & 824 & 522 & 302 & 0.6335 & 39.24\\
Encryption at rest & 16 & 768 & 90 & 678 & 0.1172 & 48.0\\
Access policy & 25 & 2,569 & 2,021 & 548 & 0.7867 & 102.76\\
Logging/Monitoring & 9 & 452 & 73 & 379 & 0.1615 & 50.22\\
IP Address binding & 4 & 428 & 302 & 126 & 0.7056 & 107.0\\
Hard-coded secret & 1 & 47 & 47 & 0 & 1.0 & 47.0\\
Outdated feature & 6 & 201 & 73 & 128 & 0.3632 & 33.5
\end{tabular}}
\label{azure_categories_results_table}
\end{center}
\end{table}

\begin{table}[htb]
\begin{center}
\caption{GCP Policy Categories Check Results}
\resizebox{\linewidth}{!}{
\begin{tabular}{c|c|c|c|c|c|c}
\textbf{Category} & \textbf{Nb of policies} & \textbf{Nb of checks} & \textbf{Nb of pass} & \textbf{Nb of fail} & \textbf{Pass rate} & \textbf{Avg nb of check/policy}\\
\hline
Admin by default & 5 & 3,275 & 3,235 & 40 & 0.9878 & 655.0\\
Encryption in transit & 3 & 792 & 28 & 764 & 0.0353 & 264.0\\
Encryption at rest & 12 & 1,522 & 210 & 1,312 & 0.1380 & 126.83\\
Access policy & 31 & 9,594 & 6,949 & 2,645 & 0.7243 & 309.48\\
Logging/Monitoring & 23 & 4,457 & 2,270 & 2,187 & 0.5093 & 193.78\\
IP Address binding & 7 & 3,256 & 2,027 & 1,229 & 0.6225 & 465.14\\
Hard-coded secret & 0 & 0 & 0 & 0 &  & \\
Outdated feature & 1 & 231 & 102 & 129 & 0.4416 & 231.0
\end{tabular}}
\label{google_categories_results_table}
\end{center}
\end{table}

\end{landscape}


\subsubsection*{\textbf{\textit{Admin by default}}}
The creation of action statements with all privileges is very rare. 
Most IAM, KMS key, and SQS policies specify which actions are allowed. 
Such results can be interpreted as a good implementation of the least privilege principle. 
The automation of infrastructure provisioning with IaC allows the creation of policies with the minimal required permissions automatically.

\subsubsection*{\textbf{\textit{Hard-coded secrets}}} 
This category and associated policy results highlight the presence of secrets in the Terraform source code tends to be rare (57 fails in 4,143 checks). 
Among those 57 fails, 17 are actual secrets, like access keys and passwords, present in the Terraform source code. 
The other 40 fail checks refer to the no usage of EKS (Elastic Kubernetes Service) Secrets Encryption, which enables containers to access secrets securely without needing to hard-code the secret in the container definition (e.g., the Dockerfile). 
Since we evaluate GitHub repositories as publicly accessible, a malicious user could use secrets to access part of the system. 
GitHub now scans repositories for known types of secrets, preventing these types of misconfiguration flaws \cite{GitHub_secrets}.

\subsubsection*{\textbf{\textit{IP Address binding \& Access policy}}}
Most resources interacting with one another are properly configured to verify the least privilege principle. 
This applies to resource-specific policies as well as IP Address binding, allowing access from all sources (0.0.0.0:0 IP address). 
Public access is often disabled, and resource policies restrict non-required authorizations. 
The key-referencing feature of IaC enables practitioners to refer to other resources in the file code easily. 
Therefore, binding resources access is easily done with IaC leaving open-access authorizations useless and vulnerable.

\subsubsection*{\textbf{\textit{Encryption in transit}}}
This category ensures all communications use secure protocols over Transport Layer Security (TLS). 
The communication has to be properly configured in the infrastructure configuration, which defines how resources communicate with one another. 
63.1\% of communication links used secure protocols, allowing for improvement.


\begin{framed}
\noindent
\textit{Access policy} and \textit{IP Address binding} categories have very successful pass rates, highlighting the benefits of using IaC tools and practitioner concern with those vulnerabilities in all evaluated cloud providers. 
For the AWS and Azure providers, we observe they present similar results in other categories when compared to GCP. 
For example, they present good implementation of \textit{Encryption in transit} and \textit{Hard-coded secrets}. 
Moreover, the GCP dataset shows relatively good adoption of \textit{Logging/Monitoring} policies, even though the individual policy results show mixed figures.
\end{framed}

\subsubsection{Commonly Neglected Security Practices}\label{rq3}

In this section, we discuss the most common security configuration flaws present in Terraform files. 
Just like in the previous section, we observe some general failing categories among all providers and further particularities. 
Next, we discuss individual aspects of each provider and then discuss in detail each of these categories.

For the AWS provider, we observe that the \textit{Encryption at rest} category policies have a pass rate of only 25.9\% (see Figure~\ref{categories_results_graph}). 
Next, the category \textit{Logging/Monitoring} presents a policy success rate of 15.4\% on 4,107 checks, while \textit{Outdated feature} closes the ranking with only 10.3\% of 1,254 checks passed.
For the Azure provider, \textit{Admin by default} privileges is implemented among 32.3\% of the checks (see Figure \ref{azure_cat_results}). 
Likewise, enabling \textit{Logs and detailed monitoring} shows low results, with a 16.2\% passing rate.
Finally, more than 63.7\% of the \textit{Outdated feature} category policies failed. 
For the GCP provider, just like in the AWS, the category (\textit{Encryption at rest} is the most neglected policy category with a pass rate of 13.8\% (see Figure~\ref{google_cat_results}). 
Next, we have the category \textit{Encryption in transit}) as the most neglected category, with a passing rate of 3.5\%. 

\subsubsection*{Worst pass rate individual policies}
Only one of the five worst pass/fail-rated policies falls outside the three poorly performing categories (see our online Appendix~\cite{thesis_replication_package}). 
Nearly all default VPC security groups do not restrict all traffic (437/440 checks failed), which we categorize as \textit{Access policy}. 
Other worst-rated policies support category findings. 
Cloud resources tend not to use encryption at rest, especially ECR repositories (1.89\%) and instances' launch configurations (5.05\%). Logging and detailed monitoring are often disabled, especially for VPCs (3.416\%). 
IMdSv1 (Instance Metadata Service v1), a depreciated service in favor of v2, is often enabled (915/963 times) \citep{maccarthaigh_2019}.

\subsubsection*{\textbf{\textit{Encryption at rest}}}
Good data encryption helps prevent leaks by third parties and ensures data integrity. 
Cloud providers now offer fully managed server-side encryption at a low cost, which eases the use of such technologies. 
Only one-quarter of resting data resources are encrypted, representing a worrying amount of vulnerable storage. 
However, our study only considers Terraform configuration and can only check for server-side encryption. 
Resting data can always be encrypted locally before uploading them to the cloud. 

\subsubsection*{\textbf{\textit{Logging/Monitoring}}}
Policies in this category check if logs and detailed monitoring are enabled on the deployed resources. 
We find log use is quite low for AWS and Azure. 
Further investigation may clarify why practitioners do not enable and use logs when deploying and maintaining cloud infrastructures, despite evidence of the usefulness of logs in system security \citep{avila2021, vaarandi}.

\subsubsection*{\textbf{\textit{Outdated feature}}}
Cloud providers often implement new service versions and security fixes automatically with no need to modify the infrastructure definition.
However, major changes may not be back-compatible, so some cloud practitioners disable automatic minor updates or stay with vulnerable versions.


\begin{framed}
\noindent
Overall, AWS and Azure cloud providers neglect the same categories, while GCP has its own particularities.
Server-side \textit{encryption at rest} is rare, despite fully managed encryption processes offered by all evaluated cloud providers. 
Hard-coded \textit{encryption in transit} is expressively neglected by GCP. 
Therefore, no related Terraform policies exist, as there is no need for practitioners to implement such practices in Terraform.
For AWS and Azure, \textit{Logging and monitoring} is neglected, despite references encouraging their use. 
Projects also become vulnerable due to \textit{outdated service versions} or by disabling automatic minor updates that may apply security fixes, especially for AWS.
On the other hand, on Azure, preventing \textit{Admin by default} privileges (by disabling admin accounts) is often neglected.
\end{framed}

\subsubsection{Differences between Cloud Providers}
\label{providers_comparison}

In this section, we explore the differences between the three studied cloud providers, AWS, Azure, and Google Cloud, and reflect on the potential reasons for the previous observations. 
First, we focus on the differences between the three datasets collected for our empirical study. 
Then we tackle the policy distributions into the eight categories defined in Section~\ref{sec:rq1}.
Finally, we compare the security policies' adoption results showcased above.

\subsubsection*{Datasets Differences} 
\label{dataset_differences}

As described in section~\ref{data_collection}, the same method for establishing the sample of GitHub repositories evaluated in this study was adopted for all evaluated cloud providers. 
In all three cases, the data collection periods lasted around one month, though they were not concurrent. 
Despite the consistency applied to the data collection method, the sample metrics may differ, which could impact results. 
Table \ref{table:datasets_metrics_comparison} lists the metrics of the three different datasets.

As discussed in section~\ref{data_collection}, the relative sizes of the AWS and Azure datasets reflect the GitHub repository population and the current usage of public cloud providers when dealing with Terraform components deploying infrastructures on those platforms \citep{richter_2022}. 
However, when we look at the GCP dataset, this assumption is not applicable.
The GitHub query ``\texttt{provider google}'' shows only 80K indexed files (compared to 417K with AWS and 101K with Azure).
This discrepancy can be explained by the popularity metrics (number of stars, forks, and contributors) that tend to be statistically higher on the repositories with GCP Terraform components (see Table \ref{table:datasets_metrics_comparison}), leading to more repositories passing the popularity filter (repositories with star count $\geq$ 2, see Section~\ref{sec:repository-filtering}).

Besides, such differences could impact the results as projects with a higher number of stars are expected to get more attention from the community, leading to a higher adoption rate of security practices, as it will be discussed in Section~\ref{sec:rq3}. 
Likewise, projects with more contributors (and forks) tend to have more diverse viewpoints and expertise, which could impact the adoption of security practices results. 
However, projects with more contributors and forks could lead to a greater spread of practice adoption and/or neglect on the platform, as it will be discussed in Section~\ref{sec:rq3}. 
On that note, we remark that the AWS dataset contains projects that seem more popular than Azure projects in terms of the number of stars but not in terms of forks and contributors.  
Finally, the larger number of projects in the AWS dataset could potentially result in a wider range of security practices being adopted. 

\subsubsection*{Taxonomies Differences} 
\label{taxonomies_differences}

AWS, Azure, and GCP provide cloud instances publicly, but the services they offer and their configuration highly differ. 
In their work, Saraswet and Tripathi (\citeyear{saraswat}) compare the characteristics of the services provided by the three cloud providers. 
Since the provided services differ, the Terraform configurations and the related security policies may also differ.

\begin{table}[htb]
\begin{center}
\caption{AWS, Azure and GCP Security Policies Taxonomy Comparison}
\begin{tabular}{c|c|c|c}
\textbf{Category} & \textbf{AWS} & \textbf{Azure} & \textbf{GCP}\\
\hline
\multirow{2}{*}{Access policy} & 29 & 25 & 31\\
& 24.0\% & 29.8\% & 37.8\%\\
\hline
\multirow{2}{*}{Admin by default} & 6 & 2 & 5\\
& 5.0\% & 2.4\% & 6.1\%\\
\hline
\multirow{2}{*}{Encryption at rest} & 35 & 16 & 12\\
& 28.9\% & 19.0\% & 14.6\%\\
\hline
\multirow{2}{*}{Encryption in transit} & 15 & 21 & 3\\
& 12.4\% & 25.0\% & 3.7\%\\
\hline
\multirow{2}{*}{Hard-coded secrets} & 4 & 1 & 0\\
& 3.3\% & 1.2\% & 0.0\%\\
\hline
\multirow{2}{*}{IP address binding} & 12 & 4 & 7\\
& 9.9\% & 4.8\% & 8.5\%\\
\hline
\multirow{2}{*}{Logging/Monitoring} & 16 & 9 & 23\\
& 13.2\% & 10.7\% & 28.0\%\\
\hline
\multirow{2}{*}{Outdated feature} & 4 & 6 & 1\\
& 3.3\% & 7.1\% & 1.2\%
\end{tabular}
\label{policy_distrib_compare}
\end{center}
\end{table}

The policies selected from the Checkov built-in catalog are security practices that can be leveraged through Terraform files. 
However, the services offered by the three cloud providers are significantly different. 
For example, AWS EC2 Instances find their equivalent in Azure Virtual Machines. 
While the two services offer the same service, their usage and configuration (as well as security specifications) highly contrasts. 
For each service, the freedom of configuration left to the user through its project Terraform component will result in different implementable security policies. 
Likewise, the default configurations will likely be different, and one cloud provider can provide a more secure service by default, leading to fewer security policies that have to be implemented by the customer. 
For instance, GCP applies most \texttt{Encryption at rest} and \texttt{Encryption in transit} best practices by default, without the need for practitioners to specify such configuration in Terraform \citep{gcp_enc_rest, gcp_enc_transit}. 
That is the motivation for selecting and mapping security policies for each cloud provider to general categories related to security best practices implemented through the provider-specific policies. 

Because policies are bonded to one cloud provider, there is no one-to-one mapping of those policies across platforms. 
Therefore, comparing each policy result between providers does not provide significant insights into security. 
For this reason, categorization can effectively address this issue and provide valuable insights. 
Instead of looking for each policy individually, we can check if the specific category of security practices is addressed or neglected in the IaC component.

Finally, the policy distribution in categories itself reveals a few insights. 
Since encryption is hard-coded into GCP services, the Encryption categories of the GCP policy taxonomy are considerably smaller than the AWS and Azure taxonomies. 
On the other hand, GCP tends to have more policies related to \textit{Logging and Monitoring} representing 28\% of the taxonomy distribution. 
The distribution of the AWS and Azure policies is more similar. 
The absence of \textit{Hard-coded secrets} policies for GCP suggests that Terraform configurations for Google Cloud do not require the use of secrets in their code, preventing this type of vulnerability from occurring. 
Overall, the AWS taxonomy contains more policies (121 vs. 84 on Azure and 82 on GCP), meaning that fewer security practices are hard-coded into the cloud service by default, leaving the implementation task to the customer through its Terraform configuration.

\subsubsection*{Adoption of Security Practices: Comparison} 
\label{results_differences}

Now we focus on comparing the results of the security policies' checking on the Terraform components of the different cloud providers under two perspectives: number of checks and pass-checking rates.

\subsubsection*{Number of Checks Comparison}

\begin{table}[htb]
\begin{center}
\caption{Check Results Metrics Comparison}
\begin{tabular}{c|c|c|c}
\textbf{Metric} & \textbf{AWS} & \textbf{Azure} & \textbf{GCP} \\
\hline
Number of GitHub projects & 401 & 137 & 274\\
Number of policies selected & 121 & 84 & 82\\
Total number of checks & 59,045 & 5,385 & 23,127\\
Median number of checks per repo & 56.0 & 17.5 & 22.0\\
Mean number of checks per repo & 160.01 & 52.81 & 108.66\\
Skew of number of checks per repo & 0.33 & 0.45 & 0.25\\
Median pass rate per repo & 71.43 & 50.0 & 57.14\\
Mean pass rate per repo & 69.11 & 51.90 & 55.92\\
Median number of checks per policy & 174.0 & 20.0 & 201.0\\
Mean number of checks per policy & 504.66 & 68.1 & 282.04\\
Skew number of checks per policy & 0.40 & 0.44 & 0.24
\end{tabular}
\label{results_metrics_comparison}
\end{center}
\end{table}

As described previously in Section \ref{dataset_differences} and \ref{taxonomies_differences}, the three datasets evaluated in this study and the service configurations targeted by the policies differ. 
As a result, the number and the nature of the policy checks produced by the Static Code Analysis also significantly differ. 
Table \ref{results_metrics_comparison} presents the reported metrics associated with the performed checks for the three datasets. 
As we can see, the Static Code Analysis of the GCP dataset executed more than ten (10) times more checks per policy than the analysis of the Azure dataset. 
For all metrics showcased in Table \ref{results_metrics_comparison}, the Azure Terraform files produced significantly fewer checks (67\% less checks per repository and 25\% less checks per policy compared to AWS). 
The lower number of checks (per repository or per policy) does not imply weaker security or weaker practice adoption. 
Instead, it reflects how often the policy can be implemented in the Terraform file, which is related to the number of cloud resources deployed and the number of policies related to those resources. 
Therefore, the low number of checks (both per repository and per policy) performed on the Azure dataset can be explained by a fewer number of resources deployed by those projects, as well as fewer security policies needing to be implemented through Terraform.

While these findings do not give insights about the security practices adoption, a lower number of checks could impact confidence in the results. 
As a result, we consider a confidence interval (with a value of 95\% confidence) related to the low number of checks into account when comparing the category's pass rate performances (see Figure~\ref{aws_azure_comparison}).
The number of checks executed on the GCP dataset stands in between the AWS and Azure ones. 
Those metrics enable us to compute the skewness of the check distribution (per repository and per policy). 
The skew metric can give an indication of whether the results are biased by a few outlier repositories and/or policies. 
A skew close to zero indicates more symmetry in the check distribution. 
Therefore, we observe that the GCP dataset results are more symmetrically distributed than the AWS and Azure datasets, which are more similar. 

Finally, we remark that the median and average pass rate per repository is higher on the AWS dataset, which can be interpreted as higher adoption of security practices by AWS practitioners. 
Once those results only showcase the adoption of practices that are leverageable by practitioners on Terraform (e.g., not taking into account the security practices hard-coded into the services by default), such figures don't imply weaker cloud security on other providers' infrastructures.

\subsubsection*{Categories Pass Rates Comparison}

\begin{figure}[htb]
\centerline{\includegraphics[width=.9\linewidth]{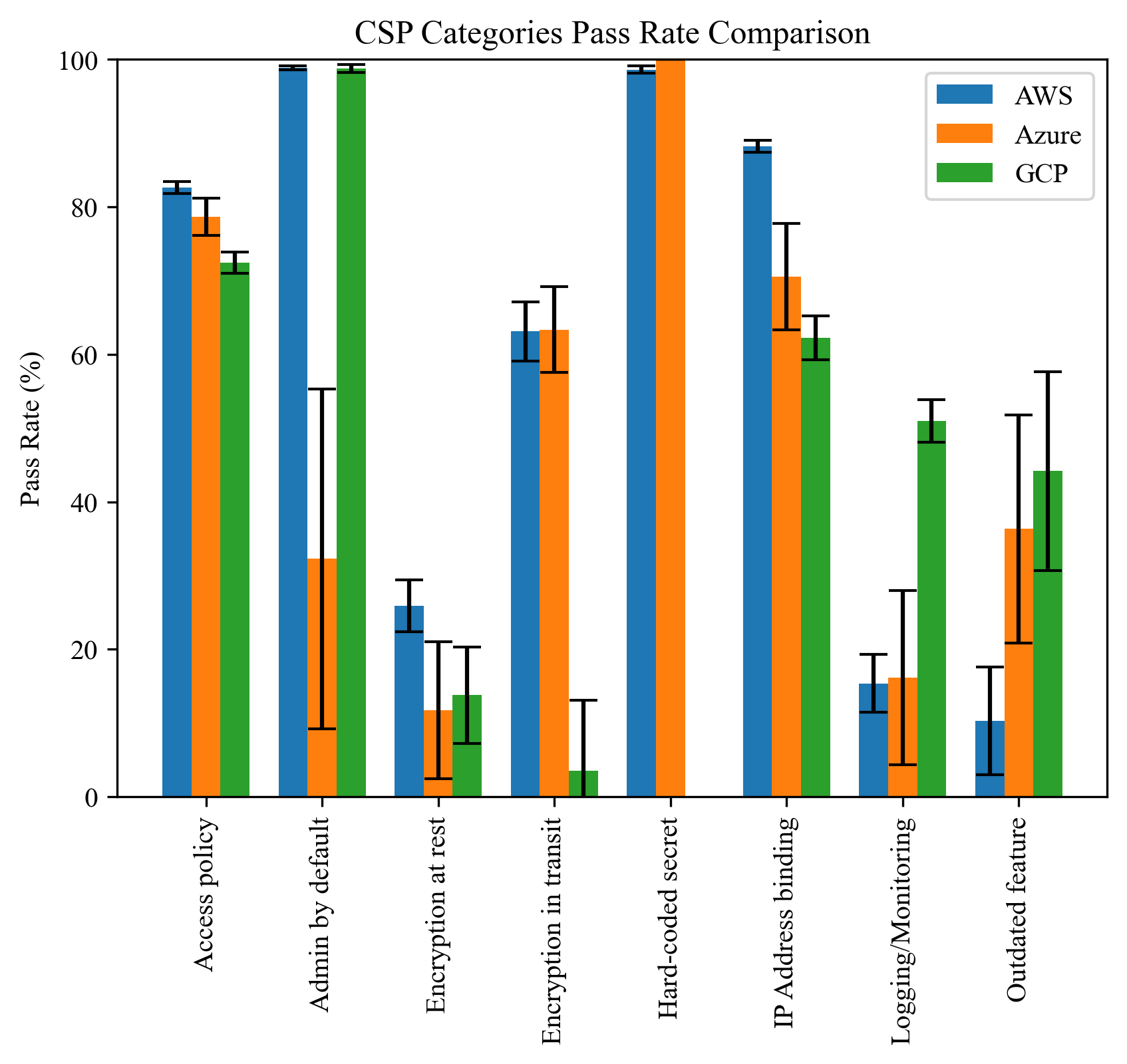}}
\caption{AWS / Azure / GCP Categories Results Comparison with 95\% Confidence Intervals}
\label{aws_azure_comparison}
\end{figure}

Now that we discussed the differences between the datasets and the policy taxonomies, we can compare the security practices adoption results between the three cloud providers. 
Figure~\ref{aws_azure_comparison} presents the categories results of AWS, Azure, and GCP. Table~\ref{tab:rq2_comparison_table} summarizes RQ2 findings, presenting the categories that tend to be well adopted or neglected for each provider and highlighting the categories that are consistently adopted or neglected across the three platforms. 

As previously discussed, the Azure dataset and policies produced significantly fewer checks. 
To better understand its statistical impact, we added 95\% confidence intervals to the categories' pass rate results. 
The confidence intervals are computed from the category pass-fail result, the desired confidence (95\%), and the total number of checks for the category. 
This helps better understand the performance difference between Azure and the other providers in the \textit{Admin by default} category. 
While the AWS taxonomy only contains 6 policies in this category, that is the category responsible for producing the most checks on the AWS dataset (20,510). 
On the other hand, the Azure \textit{Admin by default} category contains two policies, producing only 96 checks. 
While AWS uses IAM policies attached to roles and services to grant specific permissions, Azure services permissions are granted in each instance configuration. 
If the default configuration of the Azure service is secure, then there is no need for the practitioners to implement the best practice in the Terraform configuration, leading to no existing related policies, which results in the analysis producing fewer checks.

Overall, we can observe that the \textit{Access policy} category has a similar pass rate across all three providers. 
The three (3) categories \textit{Encryption in transit}, \textit{Hard-coded secrets} and \textit{Logging/Monitoring} have very close pass rates between AWS and Azure as well as \textit{Encryption at rest} between Azure and GCP. 
Figure \ref{aws_azure_comparison} shed light on the \textit{Admin by default} category, which shows a very high adoption on AWS and GCP, found on Azure with a difference of 66 percentage points. 
This finding suggests managing privileges on Azure to be significantly less convenient for Terraform practitioners, leading to neglect in this field.
Likewise, adopting \textit{Encryption in transit} policies highly differs between GCP and AWS/Azure. 
The latter has already been explained by the use of encryption by default on GCP services \cite{gcp_enc_rest, gcp_enc_transit}. 
Not using the last version of the software, as showcased by the \textit{Outdated feature} category, seems to be a more common issue on AWS.
Finally, we remark that adopting \textit{Logging/Monitoring policies} is significantly higher on GCP with a similar number of checks in this category in AWS (4,457 checks on GCP vs. 4,107 checks on AWS). 
The use of logs seems to be more common on GCP cloud infrastructures than on AWS and Azure. 
Investigating how the services of the log differ from the three cloud providers could reveal some interesting factors fostering adoption.

\begin{framed}
\noindent
Overall, the three evaluated cloud providers are significantly different. 
Their popularity (in terms of market share) leads to different dataset sizes for similar data collection duration. 
The smaller size of the Azure dataset, and consequently the lower number of checks executed, generates wider confidence intervals. 
The three cloud service providers provide similar services with different characteristics that can be configured diversely. 
As a result, the security policies that can be implemented through Terraform files also vary. 
Differences have been observed in the policy category distributions produced following the same method. 
However, some similarities have been found comparing category results, leading to insights on adopting practices like secure \textit{Access policies} and absence of \textit{Hard-coded secrets} across all three providers. 
The results also highlight how the default configuration of cloud providers impacts the adoption of security practices. 
Using encryption by default on GCP resulted in low adoption results for the few services to which it is not applied. 
Preventing \textit{Admin by default} privileges on Azure seems less convenient, resulting in low policy adoption by practitioners compared to AWS and GCP. 
All three providers show low adoption of server-side \textit{Encryption at rest} policies, despite the fully managed encryption processes offered by all cloud providers. 
Finally, a lack of \textit{Logging/Monitoring} has been observed on both AWS and Azure, but GCP seems to ease the adoption of logs-related security policies.
Those findings show that adopting security practices is actually a \textbf{shared responsibility between practitioners and providers}.
\end{framed}

\begin{landscape}
\vspace*{\fill}
\begin{table}[]
\caption{Summary of RQ2 Results for AWS, Azure, and GCP. Categories in {\color[HTML]{229954}green} are the one consistently adopted and {\color[HTML]{FF0000}red} the one consistently neglected on the three providers}
\resizebox{\textwidth}{!}{%
\begin{tabular}{l|c|c|c|c}
 & AWS                  & Azure                & Google Cloud         & Reason for observation \\
 \hline
\multicolumn{1}{c|}{\begin{tabular}[c]{@{}c@{}}Well Adopted \\Practices\end{tabular}} &
  \textit{\begin{tabular}[c]{@{}c@{}}{\color[HTML]{229954}Access policy}\\ {\color[HTML]{229954}IP Address binding}\\ Admin by default\\ Encryption in transit\end{tabular}} &
  \textit{\begin{tabular}[c]{@{}c@{}}{\color[HTML]{229954}Access policy}\\ {\color[HTML]{229954}IP Address binding}\\ Hard-coded secrets\\ Encryption in transit\end{tabular}} &
  \textit{\begin{tabular}[c]{@{}c@{}}{\color[HTML]{229954}Access policy}\\ {\color[HTML]{229954}IP Address binding}\\ Admin by default\\ Logging/Monitoring\end{tabular}} &
  \begin{tabular}[c]{@{}c@{}}Benefits from \\ Key-referencing\\  feature of IaC\end{tabular} \\
\hline
\multicolumn{1}{c|}{\begin{tabular}[c]{@{}c@{}}Often Neglected \\Practices\end{tabular}} &
  \textit{\begin{tabular}[c]{@{}c@{}}{\color[HTML]{FF0000} Encryption at rest}\\ {\color[HTML]{FF0000}Outdated feature}\\ Logging/Monitoring\end{tabular}} &
  \textit{\begin{tabular}[c]{@{}c@{}}{\color[HTML]{FF0000} Encryption at rest}\\ {\color[HTML]{FF0000}Outdated feature}\\ Logging/Monitoring\\ Admin by default\end{tabular}} &
  \textit{\begin{tabular}[c]{@{}c@{}}{\color[HTML]{FF0000} Encryption at rest}\\ {\color[HTML]{FF0000}Outdated feature}\\ Encryption in transit\end{tabular}} &
  \begin{tabular}[c]{@{}c@{}}Could be implemented \\ at another layer\\ (or hard-coded \\ in GCP services)\end{tabular}
\end{tabular}%
}

\label{tab:rq2_comparison_table}
\end{table}
\vspace*{\fill}

\vspace*{\fill}
\begin{table}[hbbp]
\begin{center}
\caption{Popularity Metrics Bins Statistics.}
\resizebox{\linewidth}{!}{
\begin{tabular}{c|c|c|c|c|c|c|c|c|c|c}
\textbf{Bin Number} & \textbf{1} & \textbf{2} & \textbf{3} & \textbf{4} & \textbf{5} & \textbf{6} & \textbf{7} & \textbf{8} & \textbf{9} & \textbf{10}\\
\hline
\textbf{Stars} Bin Limits & $[2:3[$ & $[3:5[$ & $[5:10[$ & $[10:20[$ & $[20:50[$ & $[50:100[$ & $[100:300[$ & $[300:800[$ & $[800:1500[$ & $[1,500:10,000[$\\
Repo count & 70 & 81 & 75 & 52 & 48 & 24 & 23 & 11 & 9 & 5\\
Standard Deviation & 0.0 & 0.491 & 1.387 & 2.832 & 9.258 & 13.878 & 59.214 & 163.596 & 157.331 & 3,038.093\\
Average Pass / Fail Rate & 0.689 & 0.672 & 0.718 & 0.733 & 0.829 & 0.870 & 0.780 & 0.893 & 0.983 & 0.635\\
\hline
\textbf{Forks} Bin Limits & $[0:1[$ & $[1:3[$ & $[3:5[$ & $[5:10[$ & $[10:20[$ & $[20:35[$ & $[35:50[$ & $[50:100[$ & $[100:300[$ & $[300:2200[$\\
Repo count & 79 & 84 & 57 & 51 & 43 & 22 & 17 & 17 & 17 & 13 \\
Standard Deviation & 0.0 & 0.5 & 0.493 & 1.331 & 2.789 & 4.299 & 4.190 & 15.027 & 45.338 & 492.728\\
Average Pass / Fail Rate & 0.655 & 0.709 & 0.621 & 0.687 & 0.590 & 0.889 & 0.695 & 0.574 & 0.475 & 0.673\\
\end{tabular}}
\label{bins_carac}
\end{center}
\end{table}
\vspace*{\fill}

\end{landscape}

\subsection{\textbf{RQ3: Is there a correlation between the popularity of a GitHub repository and the adoption of security best practices in its Terraform component?}}
\label{sec:rq3}

The previous research questions investigated the categorization and adoption of security practices by practitioners through the Static Code Analysis of Terraform files. 
For that, we built cloud provider bonded taxonomies that helped us compare the results between AWS, Azure, and GCP. 
So far, we have explored the repositories (code files) to derive insights on adopting security practices in Terraform files. 
However, some interesting results may be found by exploring the relation between the metadata of a repository and the adoption of practices in its Terraform component. 
Especially using metrics such as one repository's popularity in terms of the number of stars, contributors, and forks can reveal useful information regarding the good examples of security practices adopted on GitHub. 
This way, in this section, we investigate possible correlations between the popularity of a GitHub repository and the adoption of security best practices in its Terraform files.

As described in previous works by Jarczyk et al. (\citeyear{Jarczyk2014}) and Borges et al. (\citeyear{borges}), the popularity of a GitHub repository can be measured through different metrics, which are the number of stars, forks, and contributors. 
In open-source projects, those metrics tend to be intertwined. 
Many stars tend to attract new contributors, while external contributors fork the original project to make their changes and propose merge requests. 
In our case, we wonder whether popular repositories are less likely to present configuration flaws we previously identified, as more contributors might spot vulnerabilities in the Terraform files. 

To address this question, we collected the popularity metrics of our AWS dataset’s repositories and measured the overall pass rate of each repository. 
Then, for the star and fork metrics, we split the repositories into statistically distributed bins. 
Table \ref{bins_carac} presents the statistics of our bin distribution. 
The average pass rate of each bin is computed and visualized in Figures \ref{stars_graph} and \ref{forks_graph}, respectively.

\subsubsection{Number of Stars Impact}

The number of stars on a GitHub repository measures its popularity among the community. 
Most of the time, users give stars to repositories that they find interesting or useful. 
Therefore, the star count metric can be used to measure how widely a repository is used and the community's interest. 
In our dataset of open-source projects deploying cloud infrastructure with Terraform, the metrics would likely refer to the popularity of the project itself rather than the IaC component.\\

\begin{figure}[htb]
\centering
\includegraphics[width=.9\linewidth]{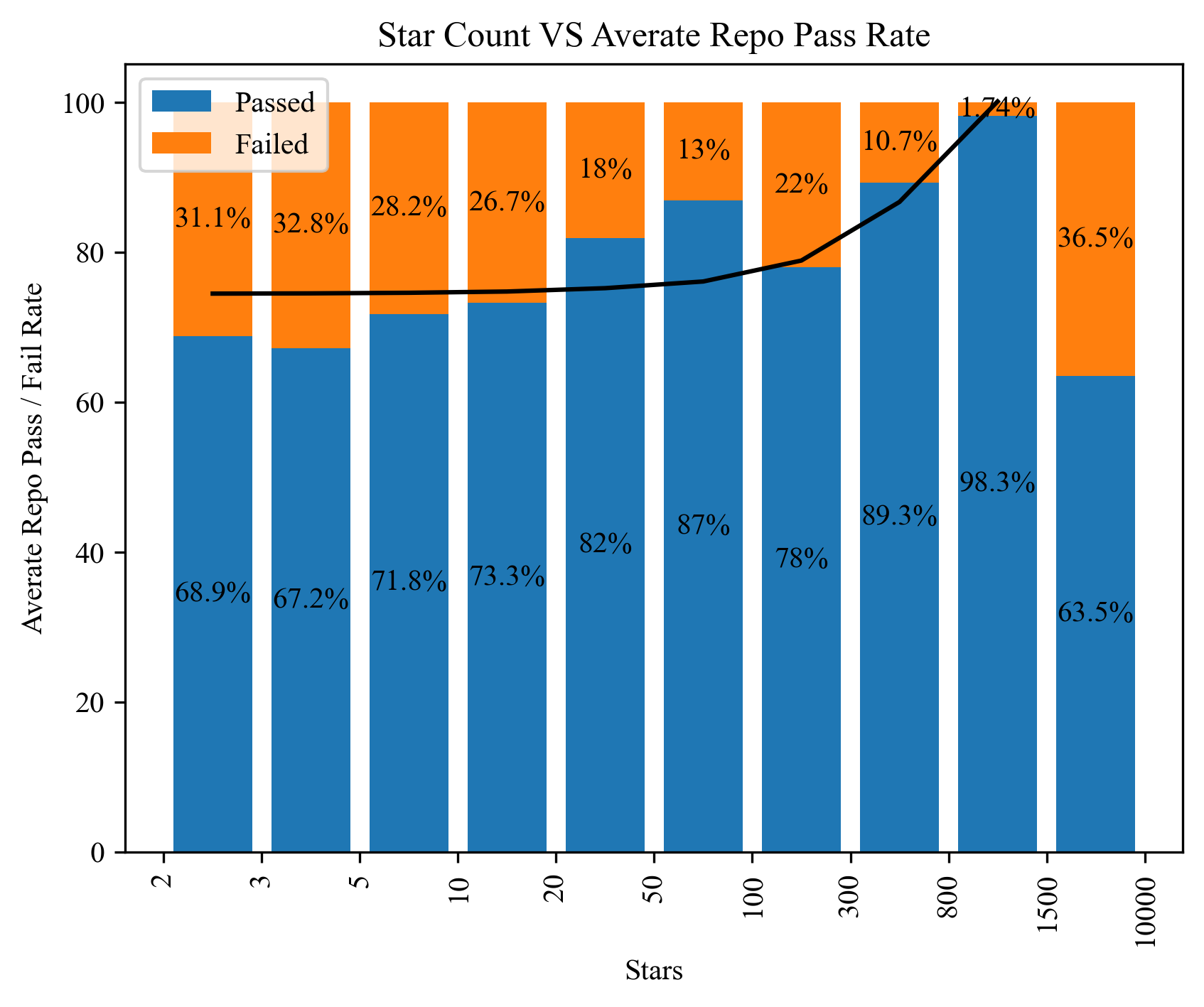}
\caption{Repo stars vs. Average Repos Pass Rate \\
Plotted Correlation: $f(stars) = 0.7446 + 0.00022*stars, R^2 = 0.692, \rho = 0.933$}
\label{stars_graph}
\end{figure}

Table \ref{bins_carac} shows the statistics of each bin, which has sorted the dataset repositories by the number of GitHub stars. 
Since projects with fewer stars are more numerous, the first bins with a short interval of low star numbers contain significantly more projects than the others. 
As a result, the first five bins tend to contain more projects while keeping a reasonably low standard deviation. 
The bin with the fewer projects contains 5 repositories, while the biggest bin contains 81. 
Figure \ref{stars_graph} visualizes the average pass rate of the repositories of each bin. 

We remark that the success rate increases with the number of stars. 
This means that the more stars a GitHub repository has, the more likely its AWS Terraform component implements the selected security policies. 
Despite a result fall with bin seven containing projects with stars between 100 and 300, the improvement seems consistent for projects with less than 1,500 stars. 
The ninth bin (projects with stars count between 800 and 1,500) has very good results with an average pass rate of 98.3\%. 
If we investigate what type of repositories appears within this bin, we find projects such as cloud secure baselines, security-focused infrastructures, and best-practice deployment demos. 
However, the last bin shows surprisingly bad results by being the one with the worst average result. 

By investigating the repositories in the last bin, we find very broad projects where the Terraform component is very small relative to the overall project. 
Cloud deployment may not be the focus of the project, which impacts the adoption of security practices. 
Since the data is not normally distributed (Shapiro test p-value of $0.00053$), we can apply the Spearman test to measure the correlation between the two metrics. 
The Spearman correlation value is $\rho = 0.933$, which indicates a significant positive correlation (p-value $= 0.00023$). 
Overall, the linear regression between the number of stars and the average pass rate has been computed, resulting in the following relation $f(stars) = 0.7446 + 0.00022*stars$. 
The function plot can be visualized in Figure \ref{stars_graph}. 

\subsubsection{Number of Forks Impact}

The number of GitHub repository forks measures how many times an original project has been copied by other users. 
A fork is a copy of a repository codebase that can be used as a starting point for one user's own project. 
A fork is often used to apply changes to the codebase before submitting it to the original repository for merge. 
This metric can be used to measure the engagement of the community through the number of people reusing the original code.

\begin{figure}[htb]
\begin{center}
\includegraphics[width=.9\linewidth]{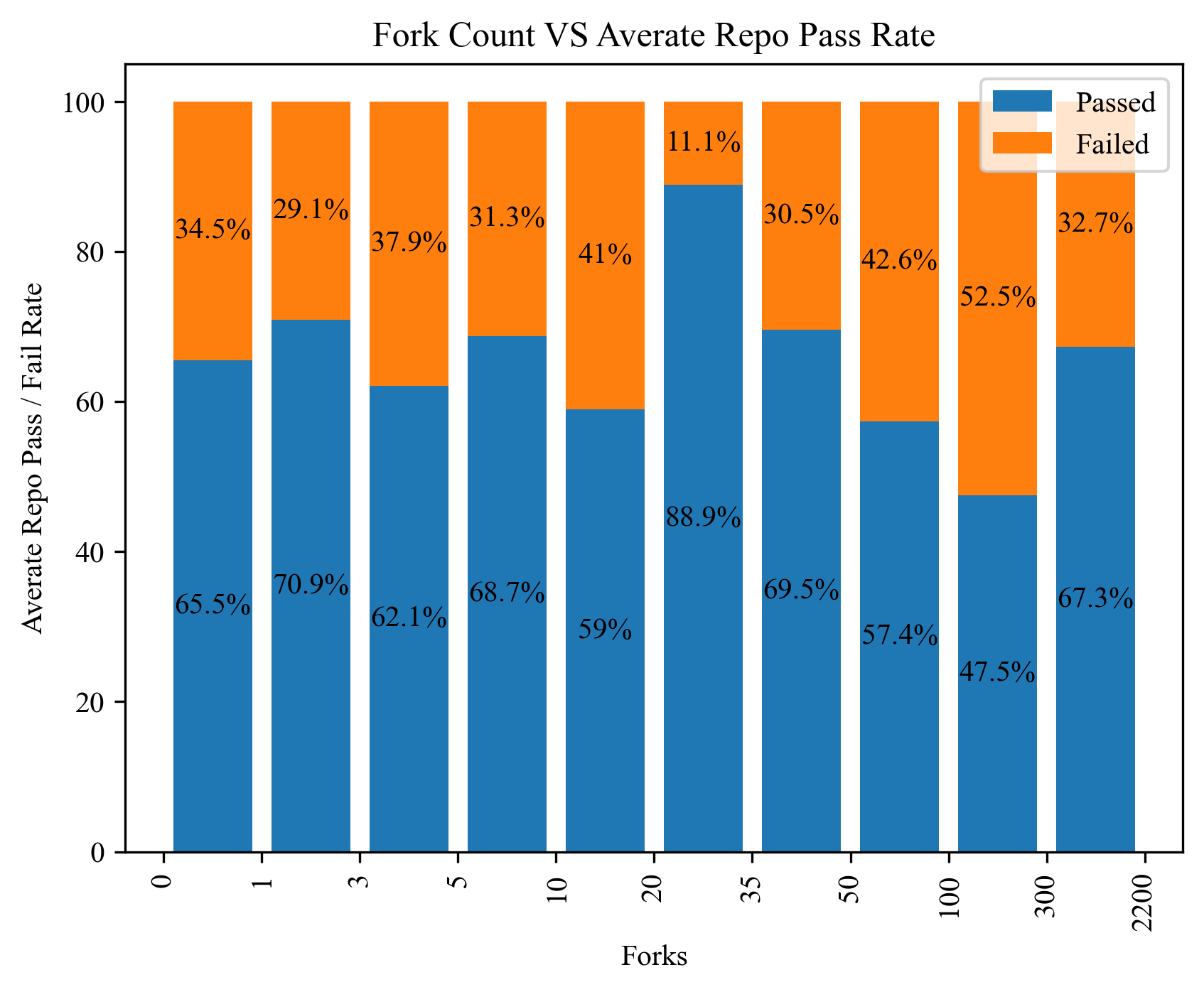}
\end{center}
\caption{Repo forks vs Average Repos Pass Rate, No Correlation Found ($\rho = -0.285$)}
\label{forks_graph}
\end{figure}

The same approach adopted for the stars investigation was done here. 
The dataset repositories have been sorted into 10 different bins. 
Table \ref{bins_carac} shows the metric distribution of each bin. 
The smallest bin contains 13 projects, while the biggest bin contains 84. 
However, as shown in Figure~\ref{forks_graph}, we could not observe any clear correlation between the number of forks and the pass rate of the SCA of the Terraform component. 
The Spearman correlation between the two variables is $\rho = -0.285$ indicating a weak negative correlation but not significant (p-value $= 0.425$). 
This means that most forked projects are not, especially the ones with good adoption of security practices. 

The absence of a clear correlation is an important finding, as by forking a project, vulnerabilities can spread across different users and subprojects. 
The worst-performing bin is the ninth one, with repositories that have been forked between 100 and 300 times. 
Such findings should raise concerns, as hundreds of projects might not implement security best practices simply because of the original project being less secure. 
This way, we encourage practitioners to check the original project infrastructure quality before forking to avoid spreading such flaws. 
Finally, it should be noted that forks can also be used to fix vulnerabilities through merge requests to the insecure original project. 
A large number of forks could showcase a more active community inclined to improve project security.

\subsubsection{Number of Contributors Impact}

The number of contributors in a GitHub repository measures how many distinct users have committed code to the project. 
A user becomes a contributor to a project as soon as it commits directly to the repository or submits a pull request that is accepted by the repository maintainer. 
The number of contributors can reflect the engagement of the community and the number of users working on the project.

\begin{figure}[htb]
\begin{center}
\includegraphics[width=.9\linewidth]{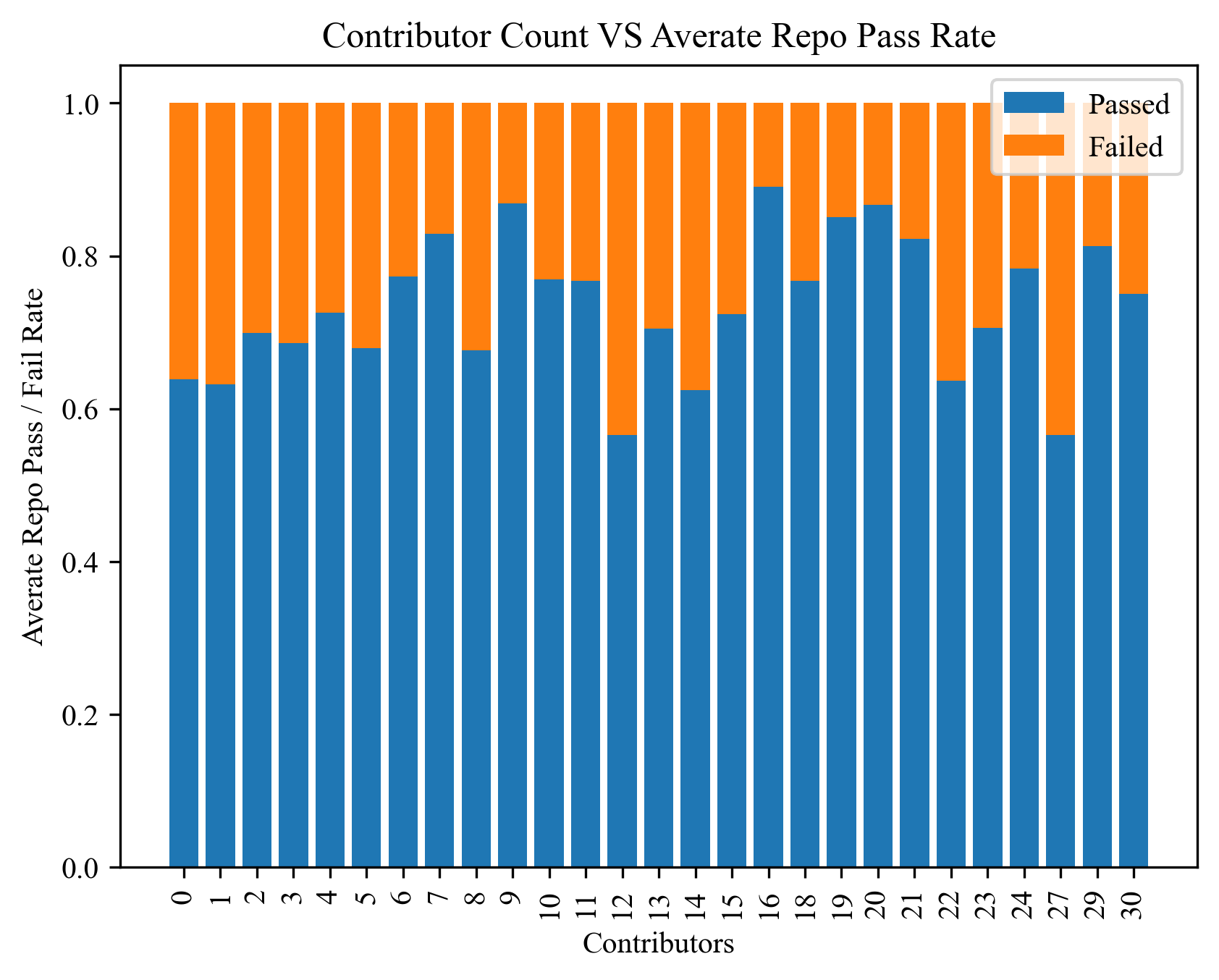}
\end{center}
\caption{Repo contributors number vs. Average Repos Pass Rate, No Correlation Found ($\rho = 0.145$)}
\label{contribs_graph}
\end{figure}

Contributors are keys in the life of open-source projects. 
A high number of contributors suppose the project takes advantage of the user’s expertise and insights. 
In our dataset, the distribution of the contributor counts is not as spread out as the two previous metrics. 
Therefore, we regrouped repositories in bins related to the precise contributor count rather than in interval bins. 
As with the results of the fork, Figure~\ref{contribs_graph} does not seem to reveal a direct correlation between the number of contributors and the adoption of security best practices in the infrastructure. 
The Spearman correlation between the two variables is $\rho = 0.145$, indicating a weak positive correlation but not significant (p-value $= 0.469$). 
Since infrastructure as code components only represent a small percentage of the total project size, we can also imagine that most contributors are focused on the application code rather than on its infrastructure.

\begin{framed}
\noindent
GitHub repository star count is positively correlated with the adoption of security practices in its AWS Terraform component. 
The more popular a repository, the most likely it is to be secure. 
Exceptions have been found for very popular projects (with more than 1,500 stars).
On the other hand, the number of forks and contributors does not seem to be correlated with the adoption of security practices. 
Projects with a high number of forks are not especially the most secure. 
Practitioners should be aware of the potential spread of vulnerable infrastructures through project forks, and be extra cautious when deploying IaC files they did not write themselves.
\end{framed}

\section{Discussion}
\label{section5}

In this section, we reflect on the previous results to suggest new guidelines for cloud practitioners and providers.
In Section \ref{section4}, we investigate GitHub repositories deploying cloud infrastructures through Terraform to measure the adoption of cloud security best practices and highlight links between GitHub metadata and infrastructure security practices adoption. 
More specifically, in Section \ref{sec:rq2}, we observe good adoption of \textit{Admin by default}, \textit{Hard-coded secrets}, \textit{IP Address binding} and \textit{Access policies} related practices in AWS infrastructures. 
Categories like \textit{Encryption at rest}, \textit{Logging/Monitoring}, \textit{Outdated features} performed significantly worse. 
We also investigated similar practices on Azure and GCP, consolidating the results of the \textit{Access policy}, \textit{Hard-coded secrets}, \textit{Encryption in transit} and \textit{Logging/Monitoring} categories. 
Finally, Section~\ref{sec:rq3} highlighted a correlation between one repository's number of stars and the average adoption of security practices. 
However, no significant relationship was observed between the number of forks and/or contributors and the adoption of security practices.

\subsection{Security Policies of Cloud Provider}

Different cloud providers offer the same type of services, but the support is packaged in different products. 
Therefore, the configuration power left to the practitioner differs from one provider to another. 
Section \ref{section4} compared the adoption of best practices between different cloud providers. 
Although the adoption results match in certain categories, we observe some differences for other categories. 
For example, differences in the way permissions are defined (\emph{IAM Policies} on AWS vs. Direct Links on Azure) and how encryption is enabled (by default on GCP) might impact how good practices are implemented through Terraform.

By investigating the policies' definitions, we can pinpoint potential reasons for the results observed in Section \ref{section4}.
\begin{itemize}
    \item Cloud providers' default configurations seem to highly impact the adoption/use of best practices. If cloud default configurations change over time, it would be possible to measure its actual impact.
    \item The number of lines of code needed to implement the practice seems to be related to its adoption. Practitioners might be more inclined to implement practices requiring only a single line of code rather than a whole new Terraform block.
    \item The key referencing feature of IaC makes easier the implementation of the least privilege principle, especially in both the \emph{Access Policies} and \emph{IP Address Binding} categories.
\end{itemize}

Extended research on the actual impact of those observations in adopting the practices could lead to more insights into the different features of IaC.

As a disclaimer, our study aims to compare the adoption of security practices by practitioners through Terraform configurations. 
Therefore, the study does not compare the actual security of the cloud infrastructure between cloud providers as this is a more complex topic to address, closely related to the cloud services design themselves.
Moreover, cloud providers constantly work on improving the security of their products. 
For instance, AWS recently changed the configuration of S3 buckets to implement some security best practices by default \cite{aws_s3_default}. 
Since the findings of this study could be impacted by future changes in cloud services designs, further future studies might be done to evaluate the current status of security policy implementation over time.

\subsection{Categories Policy Synthesis}

\begin{figure*}[htbp]
\centerline{\includegraphics[width=1\linewidth]{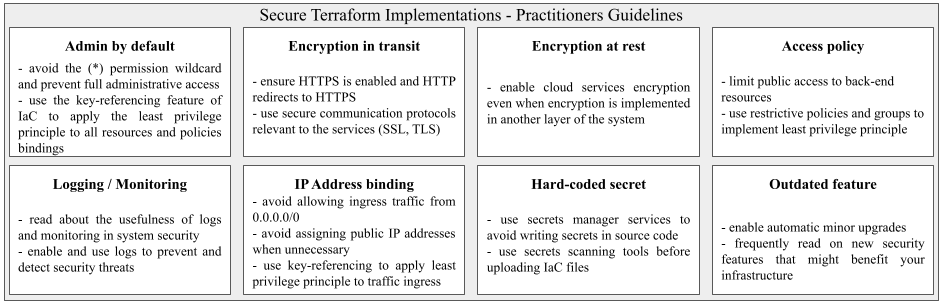}}
\caption{Practitioners Best Practices Summary For Secure Terraform Configurations}
\label{guidelines}
\end{figure*}

Analyzing the policies that address the security best practice standards and their adoption through the Static Code Analysis of Terraform files of real-world projects reveals key principles when designing secure infrastructures. 
Figure \ref{guidelines} summarizes the guidelines for each category reported in Section \ref{section4}.

To provide guidelines to cloud practitioners, we synthesize the policies contained in each category and further extend them to the potential benefits and impact of implementing them.

\textbf{\textit{Admin by default:}} Avoiding wildcard permission and applying the least privilege principle via the key-referencing feature of IaC helps ensure that minimal required access is granted to the right services, preventing unauthorized access and ensuring resources can not be misused by potential attackers.

\textbf{\textit{Encryption in transit:}}
Enabling HTTPS ensures that all communication between the client and the server is secure and encrypted. 
This practice helps prevent eavesdropping and tampering of data in transit. 
Using secure communication protocols such as SSL and TLS further enhances security by providing a secure channel for data transfer and authentication. 

\textbf{\textit{Encryption at rest:}}
Encrypting all data at rest ensures that sensitive information is protected even during a data breach. 
Enabling encryption provided by cloud services gives an extra layer of security with no additional cost; this practice helps to protect the data even if another layer is compromised.

\textbf{\textit{Access policy:}}
Limiting public access to back-end resources is crucial in preventing unauthorized access and potential breaches. 
Following the least privilege principle by applying IAM policies (on AWS) and groups ensure that resources only have access to services and permissions needed to perform specific tasks. 

\textbf{\textit{Logging / Monitoring:}}
Monitoring logs can provide valuable insights into system activity, helping to identify patterns that may indicate security breaches \citep{avila2021, vaarandi}.

\textbf{\textit{IP Address binding:}}
Avoiding ingress traffic from 0.0.0.0/0 and not assigning public IPs when unnecessary helps to limit the attack surface and reduce the risk of unauthorized access. Ensuring that only necessary traffic is allowed by using key-referencing in IaC allows for the implementation of the least privilege principle and limits potential vulnerabilities.

\textbf{\textit{Hard-coded secrets:}}
Using secret manager services ensures that sensitive information, such as passwords and API keys, are securely stored and not accidentally committed to source code. 
As an additional measure, using secrets scanning tools before uploading IaC files also helps to identify and remove any potential secrets in the code before it is deployed \citep{GitHub_secrets}.

\textbf{\textit{Outdated feature:}}
Enabling automatic upgrades ensures that the infrastructure is always updated with the latest security patches, addressing known vulnerabilities and reducing the risk of data breaches.

\subsection{The Use of GitHub}

In Section \ref{rq3}, we gathered GitHub repositories popularity metrics and looked at potential correlations with adopting security practices. This way, we discuss two potential insights practitioners might be aware of:
\begin{itemize}
    \item The star number metric tends to correlate with the adoption of security practices in the project cloud infrastructure. The more stars a project has, the more likely it is to adopt security best practices.
    \item No correlation can be established between the number of forks or contributors metrics with the adoption of infrastructure security best practices. We suggest practitioners be cautious when reusing existing Terraform files to prevent the propagation of infrastructure vulnerabilities.
\end{itemize}

\textbf{Overall Guideline:}
All practitioners must keep up to date with security best practices and standards to be able to proactively improve the security of their infrastructures. 
We recommend integrating source code scanning tools in the development pipeline to automate checks for IaC misconfigurations. 
This helps to ensure infrastructure code adheres to established practices and reduces vulnerabilities from early misconfigurations.\\


\subsection{Policy Set Completeness}

In Section \ref{sec:rq1}, we carefully identified which security guidelines have matching Terraform implementations policies that can be directly mapped to recognized industry standards. 
Even though the process aims to build a qualitative set of recognized security policies to ensure the relevance of the scanning results, not all cloud security practices might have been identified and selected. 
Therefore, our policy set could be limited in two ways, which we discuss now.

\subsubsection{Unsupported Security Guidelines on Terraform}

Some security guidelines specified in cloud regulations can not be implemented through Terraform for several reasons.
For example, some guidelines related to account configuration rather than the infrastructure definition itself, or those the required feature is not supported by the interface provided by Terraform.

The CIS standards specify whether one guideline can be automatically implemented or if it requires a manual action. 
One of such policies is the \textit{CIS AWS Foundation Benchmark} specification \textit{\textbf{1.18 Ensure IAM instance roles are used for AWS resource access from instances}}. 
In fact, this specification requires logging to the AWS web console with the appropriate permissions to view IAM Account Settings to ensure the correct roles are used. 
This specification can not be implemented through Terraform and therefore, no related scanning policy exists.

\subsubsection{Unsupported Security Practices in Regulations and/or Industry Standards}

On the other hand, practitioners can implement security practices that are not contained in industry standards and/or cloud regulations. 
Investigating such practices can generate interesting findings to understand why practitioners implement security practices that does not come from compliance requirements.

For example, the set of built-in policies in checkov contains the following policy: 
\textit{\textbf{Ensure that auto Scaling groups that are associated with a load balancer, are using Elastic Load Balancing health checks}}.
This policy could not be mapped to either one of the two industry standards, suggesting that no cloud regulations require the adoption of such practice. 
However, using health checks ensure that the instance is up and ready to perform its task before receiving traffic from a load balancer. 
This can be seen as a security practice, ensuring that no unprepared (and potentially vulnerable) instance receives external requests.

As we saw through those two examples, the set of policies extracted from industry standards used to scan the IaC components of our datasets might not contain all relevant security policies. 
Investigating the gap between standards and actual practices could lead to interesting findings. 
Practices not contained in standards but commonly implemented by practitioners could suggest new revisions of cloud regulations to implement more recent and more secure practices. 
Likewise, studying the security practices that can not be implemented through Terraform could lead to qualitative findings on the practitioners' security habits that can not be empirically measured with code artifacts. 
Such paths could lead to insightful future research in the field.

\section{Threats to Validity}
\label{section6}
Our empirical analyses and evaluations naturally leave open
a set of potential threats to validity, which we explain in this section.

\textbf{Construct Validity:}
We acknowledge the occurrence of potential methodological threats. 
For establishing our sample of GitHub projects, we used cloud provider definitions to Terraform file code snippets, although IaC components form a small percentage of a project's total size.
For that, we had to look for cloud provider imports on Terraform files.
We know there are more common ways of collecting GitHub repositories based on projects' titles, descriptions, associated programming languages, or topics.
However, these approaches would not have worked for us, as general GitHub Search does not track content code file information.

In the same way, due to GitHub API limitations, the data collected is linked to the data collection periods. 
To mitigate this risk, we adopted rounds of data collection periods that lasted around a month for each cloud provider. 
By the end, we observed that after each new round, the number of new projects added to the datasets was becoming low, showing that most of the active projects with Terraform files may have been collected for each cloud provider. 

On top of that, even though the collection period lasted around one month for all three datasets, the collections have not been executed simultaneously, impacting the ability to compare the results between providers. 
To mitigate this risk, we reduced the overall extent of the collection period to less than 6 months. 
Furthermore, we understand that popularity might not be a relevant metric to filter GitHub repositories. 
This filter helped reduce the number of repositories to analyze manually.
The popularity threshold has been set quite low ($>1$ star) to mitigate the risk of bias in the dataset. 

The mapping of policies into their associated categories was a manual process conducted by one researcher.
Eventually, some misjudgments may have happened, considering this process was based on one single author's judgment.
In order to address this threat, external consultants were individually asked to evaluate the proposed mapping.
Although we have reported a high agreement among the consultants regarding the mapping, we acknowledge that providing the mapping could have influenced the consultants' analyses to some extent.
However, given the self-explainable nature of the policies and the distinctiveness of the categories, we believe new mapping attempts to be straightforward, resulting in no or small differences from our current proposed mapping.

Finally, to perform the infrastructure scanning and vulnerability detection, as well as predefined security policies, we used the open-source tool Checkov.
We believe possible risks associated with inaccurate analysis might be mitigated by its auditability and open development community.

\textbf{Internal Validity:}
Although we have adopted and selected standard industry-recognized cloud security and privacy practices, we may not have identified all of them at the infrastructure level. 
We selected only Terraform secure configuration checks implemented by Checkov policies that map to a specification of either the CIS Amazon Web Services Foundations (v1.4.0) \citep{cis} or AWS Foundational Security Best Practices (v1.0.0) \citep{aws_fundamentals}, the CIS Microsoft Azure Foundations Benchmark (v2.0.0) \citep{cis_azure} and CIS Google Cloud Platform Foundation Benchmark (v1.3.0).
This way, it is expected that other security best practices might exist. 
To address this threat, we consider recent versions of these standard guidelines that might reflect the current status of these policies' adoption/usage by the community. 

In the same way, it is possible that we miss policies in our study commonly adopted by the community.
Although the open-source nature of Checkov mitigates the risk of omission as contributors may regularly add new policies, we would still miss these policies considering they are not present in these standard guidelines yet.  
Likewise, our reliance on only four industry standards is mitigated by their broad acceptance~\citep{stultienscompliant}. 
Finally, we categorize the investigated policies by importing categories from previous work~\citep{rahman} and also proposing new ones. 
Although this categorization process has been conducted by one single researcher, we validated our mappings with independent cloud security experts, showing a high agreement with our proposed categorization.

\textbf{External Validity:}
We acknowledge other external considerations might represent a validity threat. 
GitHub, the source of our dataset, hosts publicly accessible, often open-source repositories and, therefore, might not fully represent the industry context in which projects are often closed-sourced. 
Since security may not be a priority for some projects, we addressed this threat by manually removing projects declared purposefully insecure.
However, for the remaining projects, we can not guarantee that practitioners always aim to maximize security. 
Still, regarding our sample, although we have selected repositories with different sizes, programming languages, and domains, our results might not reflect the reality of projects out of these properties.

Also, as mentioned above, different security practices can be leveraged at different levels. 
For example, weak server-side encryption can be mitigated by users who encrypt their data locally before sending it to the cloud. 
As a result, it is expected that encryption-related policies represent, at most, a lower bound of these policies' implementation. 
Finally, as with any scanning software, the results are likely to contain false negatives and positives. 
Due to the popularity and high usage of Checkov by the community, we assume the tool is accurate enough.
So that the inaccurate analysis might not significantly impact and bias the aggregated results.
Regarding the applicability of our proposed categorization, we believe other static analysis tools could benefit from our findings in order to run new studies.
For that, only new scripts might be required to check the policies' implementation and the final report, as the other scripts might be reused from this study.

\section{Related Work}
\label{section7}
In this Section, we summarize and discuss some related work. First, we discuss some studies regarding IaC, focusing on maintainability.
Next, we discuss current standards and benchmarks for security policies.
Finally, we discuss some studies regarding the usage of static analysis tools for IaC.

\subsection{Infrastructure as Code \& Terraform}

Few empirical studies have investigated IaC code maintainability. 
Bent et al. (\citeyear{bent}) and Schwarz et al. (\citeyear{schwarz}) have examined \emph{Puppet} and \emph{Chef} configurations. 
A systematic mapping study by Rahman et al. (\citeyear{RAHMAN201965}) highlighted the need for research into quality issues such as security flaws in IaC scripts. 
In further studies, the authors have also worked on characterizing configuration defects \citep{rahmanwill}, identifying seven common security flaws in Puppet scripts through qualitative analysis \citep{rahman}. 
However, those studies focus on system configurations and installations and are not specific to the infrastructure layer. 

The closest work to our study has been done by Iosif et al. (\citeyear{iosif2022large}). 
As we do here, the authors performed an empirical study investigating security vulnerabilities in 8256 GitHub repositories of Terraform files but only deploying AWS infrastructures.
Next, they used three scanning tools to check for security vulnerabilities. 
Due to their collection method, their dataset only contains repositories where HCL (Terraform language) is the most used language of the project (e.g., $>$ 50\% of the project size), whereas Jiang et al. (\citeyear{jiang}) suggested that in real-world projects, IaC components co-exist with other types of files with a median around 11\% of the total project file number. 
As a result, they focused on malpractices based on the tools' full lists of built-in policies without investigating the industrial relevance of the policies.

Unlike the previous related work, in this study, we focus on a smaller dataset of 812 GitHub repositories from three different cloud providers (AWS, Azure, and GCP), but with recent activity by checking the last commits dates, where the IaC component only represents a small percentage of the project size. 
Moreover, we identify and categorize industry-recognized security best practices while investigating the adopted and neglected practices.
Therefore, despite the growing prevalence of projects containing Terraform configuration files on hosting platforms like GitHub, we know of no empirical studies besides our own investigating the adoption of security practices (good or bad) at the infrastructure level in \emph{real-world} projects.

\subsection{Cloud Security \& AWS Security Best Practices Standards} \label{security_standards}
To help practitioners comply with cloud infrastructure security and privacy regulations, providers and independent organizations have created sets of best practices and security benchmarks. 
Some form the basis of security certifications, of both cloud deployment audits and practitioners' knowledge. 
These help build customer trust and have become industry standards for secure cloud infrastructure \citep{auditing, sec_engin}. 
The two main standards for AWS are the Center for Internet Security (CIS) AWS Foundations Benchmark \citep{cis} and the AWS Foundational Security Best Practices standard \citep{aws_fundamentals}. 
According to Stultiens (\citeyear{stultienscompliant}), those two standards are widely respected and adopted.
In the same way, the CIS also developed security best practices standards on Azure \citep{cis_azure} and Google Cloud \citep{cis_google}.
In this study, we used the proposed standard practices to check their implementation on a sample of repositories.
Although our goal is not to evaluate the applicability and coverage of these practices, we discuss possible fields that could be added to these benchmarks.


\subsection{IaC Static Code Analysis (SCA) Tools} \label{sca_tools}
Guerriero et al. (\citeyear{guerriero2019adoption}) investigate the usage of IaC by practitioners, the current support for IaC, and the current needs of the community. 
For that, the authors performed 44 semistructured interviews with practitioners from different roles and companies. 
As a result, they highlight the difficulty of testing and maintaining IaC code. 
However, with the rise of IaC, several code-scanning tools have been released to help practitioners and automate tests in integration pipelines. 
For example, \emph{tfsec, terrascan, semgrep, checkov} are all able to scan Terraform code files and look for security misconfiguration \cite{tfsec, terrascan, semgrep, bridgecrewio}. 
These tools provide built-in misconfiguration catalogs and can be implemented in delivery pipelines. 
Among those tools, \emph{checkov} provides the larger supported compatibility with other IaC tools and cloud providers. 
Furthermore, it uses a graph-based scanning approach enabling complex security tests with deployment context awareness (e.g., tests regarding several resources simultaneously).

No prior work focuses on the infrastructure provisioning and management stage of IaC in real-world projects.
In our evaluated sample of projects, the IaC component co-exists with other types of code, which excludes Terraform templates and tutorials not deploying actual applications. 
Closely related to the cloud provider platform and the practitioners building cloud architectures, this layer often involves Terraform configuration files. 
SCA tools compatible with Terraform exist and could be used to build relevant empirical studies. 
Furthermore, we evaluate the applicability of \emph{checkov} on three different cloud providers (AWS, Azure, and GCP). 
No prior study compares the adoption of security practices between public cloud providers with datasets of these characteristics.


\section{Conclusions}
\label{conclusion}
Infrastructure as Code provisioning tools are very useful but do not automatically preclude misconfiguration and security risks. 
This work analyzes 287 secure Terraform configuration snippets from different cloud providers (AWS, Azure, and GCP), categorizing them into eight groups and empirically investigating the prevalence of their use in 812 recently active open-source GitHub repositories. 
For that, we first selected and categorized standard industry-recognized security policies. 
Then, we mined and filtered GitHub repositories to ensure the dataset consisted of genuine projects being deployed on specific cloud providers through Terraform. 
Next, we used checkov, a static code analysis tool, to scan these projects for security vulnerabilities. 

As a result, we found that some security policy categories, such as \textit{Access policy} and \textit{IP Address binding}, tended to be consistently applied by practitioners in all evaluated cloud providers. 
We also observe that AWS and Azure providers present similar results in other categories when compared to GCP.
For example, they present good implementation of \textit{Encryption in transit} and \textit{Hard-coded secret}. 
When it comes to the neglected categories, we observe that \textit{Encryption at rest} is the lowest implemented category. 
Once again, we observe AWS and Azure cloud providers neglect some same categories while
GCP has its own particularities. 
For example, AWS and Azure neglect \textit{Logging/Monitoring} and \textit{Outdated feature}, while GCP neglects \textit{Encryption in transit} and \textit{Hard-coded secret}.
Based on these findings, we provide guidelines that cloud practitioners could adopt to enhance the security of their IaC code. 
Finally, regarding GitHub measures correlated with best practice adoption, we observe a positive, strong correlation between a repository number of stars and adopting practices in its cloud infrastructure.

Our findings indicate that IaC configuration poses a major risk and confirm that, despite the availability of security scanning tools, there is a lack of adoption of best practices in open-source projects. 
More effort is required to raise awareness about the significance of applying IaC best practices in open-source projects.
In the same way, we believe these practices would become reachable easier by making IaC security scanning tools more widely available and accessible to developers as part of frameworks for continuous delivery and continuous deployment. 
In addition, efforts should be made to give education and resources on how to use these technologies successfully and to encourage cloud application developers to prioritize security in their deployment processes. 

In conclusion, Infrastructure as Code is a very powerful tool benefiting from the upsides to writing code like collaboration, code analysis, or pipeline integration. 
Building secure cloud infrastructures can be challenging for practitioners. 
The adoption of security policies in Terraform components still varies according to the best practice category across different cloud providers. 
We hope this study might support cloud practitioners and providers to improve towards more secure infrastructures through valuable insights on state-of-the-art practices.

\section{Acknowledgements}
\label {acknowledgements}
This work is partly funded by the Natural Sciences and Engineering Research Council of Canada (NSERC), Fonds de recherche du Québec (FRQ), and the Canadian Institute for Advanced Research (CIFAR). 

%

\bibliographystyle{plainnat}      
\bibliography{emse_paper}   

%
%

\end{document}